\documentclass[
 reprint,
superscriptaddress,
 amsmath,amssymb,
 aps,
 prx,
]{revtex4-1}

\usepackage{graphicx}% Include figure~files
\usepackage{dcolumn}% Align table columns on decimal point
\usepackage{bm}% bold math
\usepackage{braket}
\usepackage{color}

\makeatletter
\def\convertto#1#2{\strip@pt\dimexpr #2*65536/\number\dimexpr 1#1}
\makeatother

\begin{document}

\title{An on-chip architecture for self-homodyned nonclassical light}

\author{Kevin A. Fischer}
\email{kevinf@stanford.edu}
\affiliation{E. L. Ginzton Laboratory, Stanford University, Stanford CA 94305, USA}
\author{Yousif A. Kelaita}
\affiliation{E. L. Ginzton Laboratory, Stanford University, Stanford CA 94305, USA}
\author{Neil V. Sapra}
\affiliation{E. L. Ginzton Laboratory, Stanford University, Stanford CA 94305, USA}
\author{Constantin Dory}
\affiliation{E. L. Ginzton Laboratory, Stanford University, Stanford CA 94305, USA}
\author{Konstantinos G. Lagoudakis}
\affiliation{E. L. Ginzton Laboratory, Stanford University, Stanford CA 94305, USA}
\author{Kai M\"uller}
\affiliation{Walter Schottky Institut, Technische Universit\"at M\"unchen, 85748 Garching bei M\"unchen, Germany}
\author{Jelena Vu\v{c}kovi\'c}
\affiliation{E. L. Ginzton Laboratory, Stanford University, Stanford CA 94305, USA}

\date{\today}% It is always \today, today,

\begin{abstract}
In the last decade, there has been remarkable progress on the practical integration of on-chip quantum photonic devices yet quantum state generators remain an outstanding challenge. Simultaneously, the quantum-dot photonic-crystal-resonator platform has demonstrated a versatility for creating nonclassical light with tunable quantum statistics, thanks to a newly discovered self-homodyning interferometric effect that preferentially selects the quantum light over the classical light when using an optimally tuned Fano resonance. In this work, we propose a general structure for the cavity quantum electrodynamical generation of quantum states from a waveguide-integrated version of the quantum-dot photonic-crystal-resonator platform, which is specifically tailored for preferential quantum state transmission. We support our results with rigorous Finite-Difference Time-Domain and quantum optical simulations, and show how our proposed device can serve as a robust generator of highly pure single- and even multi-photon states.
\end{abstract}

\pacs{Valid PACS appear here}% PACS, the Physics and Astronomy
\maketitle

\section{Introduction}

The photonic integrated platform has been explored extensively for the implementation of quantum technologies~\cite{OBrien2009-dq}. Most of the components in the all-optical quantum circuits are relatively efficient classical devices that can already be ordered from a foundry~\cite{Soref2006-ae}, with the exception of one critical technology that can still benefit from improvement: the integrated quantum state generator. While some on-chip approaches look to multiplex many low efficiency sources together~\cite{Collins2013-qc,Galli2014-bs,Meany2014-uu}, in our work we look to design a single highly efficient integrated source~\cite{Aharonovich2016-pq}.

A promising route towards generating quantum states of light on a chip exists in the quantum-dot (QD) photonic-crystal-cavity (PCC) architecture~\cite{Englund2011-lx,Arcari2014-ch,LPOR201500321,lodahl}. Here, the quantum dot is a solid-state artificial atom (e.g. an island of InAs embedded within a GaAs host matrix) that yields the strong quantum non-linearity required for nonclassical or quantum state generation. The quantum dot interacts with a photonic crystal cavity~\cite{Akahane2003-ma} that has been shown to couple efficiently to an adjacent integrated waveguide~\cite{Waks2007-lu}. As a whole, the system enables extremely large enhancement of the light-matter interaction rate and can lead to a hybridization of light and matter quantum states in the strong coupling regime~\cite{Yoshie2004-yd,Englund2007-wy,Reithmaier2004-fu}, known as the Jaynes-Cummings ladder~\cite{Shore1993-zy}. Due to the formation of these new quantum states, known as polaritons~\cite{Laussy2012-hr}, QD PCC systems can generate not just single photons~\cite{Muller2015-il,Muller2015-om,Muller2016-fs} but potentially multi-photon wavepackets that may be more efficient light sources in integrated quantum circuits~\cite{Sanchez_Munoz2014-zk,Dory2016-ca}. However, large dissipation rates in such platforms have previously impeded the generation of pure quantum states of light.

Recently, a self-homodyning interferometric technique was suggested as a promising route to overcoming these intrinsic dissipations~\cite{Fischer2016-dj}. In this scheme, light is scattered both through the cavity-dot system and a secondary channel in a way that effectively isolates the interesting quantum emission from the system. For the original experiments, the secondary channel was formed by the continuum modes of the photonic crystal cavity structure that exist above the light line, potentially arising from band-edge modes and higher-order cavity resonances. When utilizing self-homodyne interference in this off-chip configuration, single-photon generation of high purity was demonstrated~\cite{Muller2016-fs} and exciting initial results suggest the observation of multi-photon states~\cite{Dory2016-ca}. Unfortunately, precise control of both the amplitude and phase for this interference is experimentally challenging and a more reproducible method is desired.

By instead utilizing the continuum modes of a photonic crystal waveguide for the secondary channel, the self-homodyne technique could be improved in experimental precision---meanwhile its integration on chip might enable its use as a quantum light source in a photonic integrated circuit. In this paper, we propose such a fully integrated and fabricable structure. To analyze its behavior, we perform rigorous Finite-Difference Time-Domain~\cite{Waks2007-lu} (FDTD) and quantum optical simulations~\cite{Johansson2013-cv} (we have included a Jupyter notebook detailing the quantum optical calculations in the arXiv submission).  Furthermore, we delve more deeply into the underlying mechanics of the self-homodyne interference technique and justify how it allows for strongly enhanced single- and multi-photon emission even in the presence of highly dissipative cavities.

\begin{figure}
  \includegraphics[width=\columnwidth]{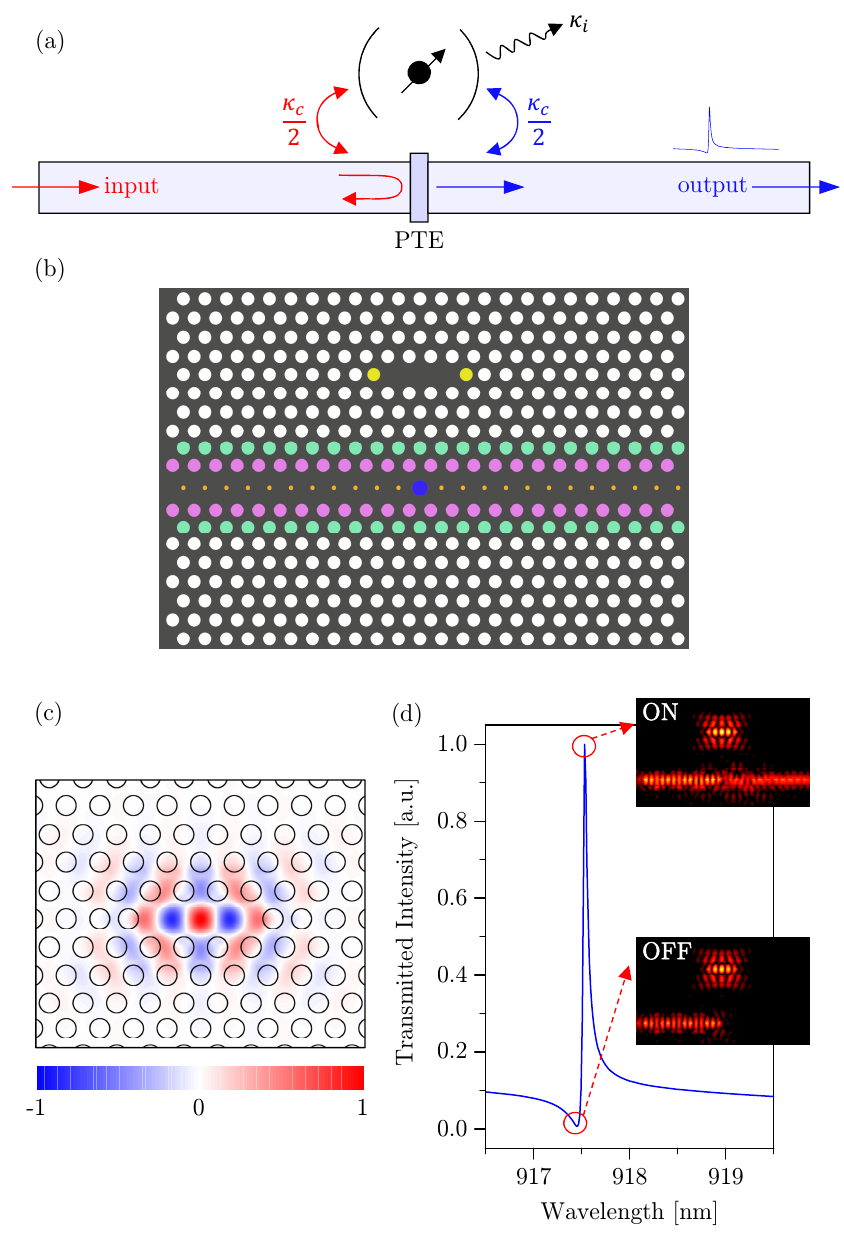}
  \caption{(a) Schematic representation of our proposed integrated version of the quantum-dot (QD) photonic-crystal-cavity (PCC) platform. The output light is scattered through both the Jaynes-Cummings system and through a partially transmitting element (PTE). (b) Proposed dielectric structure for preferentially selecting for the quantum-mechanically scattered light in transmission through the photonic crystal waveguide. (c) Field profile of the L3 photonic crystal cavity's high Q mode, calculated with Finite-Difference Time-Domain (FDTD) simulations. (d) Fano-resonant lineshape in transmission, arising from the scattering through two different channels (with lattice constant $a=241\,\textrm{nm}$). The dip in the lineshape is where the optimal self-homodyne interference occurs and the quantum light is emphasized. Inset shows FDTD mode distributions in steady state at the point of optimal self-homodyne interference and at the point of maximum transmission (wavguide fields amplified for clarity).}
  \label{figure:1}
\end{figure}

% ------------------------------------------ section ------------------------------------------ 
%%%%%%%%%%%%%%%%%%%%%%%%%%%%%%%%%%%%

\section{Proposed device}

We begin with a discussion of the classical transmission behavior of a waveguide-coupled PCC platform, without simulating the embedded quantum dot. A schematic illustration of our proposed device is shown in figure~\ref{figure:1}a. In this configuration light is injected through a waveguide, which then scatters into the Jaynes-Cummings system and out to the output section of the waveguide with rates $\kappa_c/2$---this scattering path is known as the discrete channel owing to its single mode character. Scattering through the discrete channel alone would result in a symmetric Lorentzian resonance in transmission. Additionally, light may scatter directly from the input channel to the output channel through a partially transmitting element~\cite{Yu2014-qp} (PTE)---this scattering path is known as the continuum channel owing to its flat frequency response. The interference between these two scattering channels then results in an asymmetric Fano-resonant lineshape~\cite{Fano1961-oa} in transmission. This type of device has been explored classically~\cite{Yu2014-qp,Yu2016-tn,Yu2015-mh} and semi-classically~\cite{Zhao2016-qs,Li2016-sm,Yu2016-me} for switching and lasing applications. However, in our paper we will explore this device from a fully cavity quantum electrodynamical (CQED) perspective~\cite{Haroche2006-cq} by studying the quantum statistics of the transmitted light~\cite{Loudon2000-li}. As we will see, an optimally tuned Fano interference will allow the system to act as an excellent nonclassical state generator.

Next, we discuss our proposed dielectric structure for optimal exploration of CQED effects (figure~\ref{figure:1}b). This structure is based on a triangular photonic crystal of radius ${r_0=0.3d}$ and thickness $t=0.6d$ (where $d$ is the lattice constant) and has a number of significant features:
\begin{itemize}
\item An L3 defect was chosen for the photonic crystal such that it is optimized for ease of coupling to quantum emitters, but may still readily reach the strong coupling regime~\cite{Yoshie2004-yd,Englund2007-wy}. The mode profile of the target resonance is shown in figure~\ref{figure:1}c, which is a standard mode profile for an L3's high Q mode~\cite{Akahane2003-ma}---a quantum dot may strongly couple to the mode over a relatively large locus of positions.
\item The photonic crystal waveguide (figure~\ref{figure:1}b) relies primarily on a single TE mode~\cite{Joannopoulos2011-ax} such that stray quantum emitters within $40\,\textrm{nm}$ of the holes (if randomly positioned) will be interrupted by the waveguide's line of defects, with small holes of radius $r=0.099d$ (orange). The presence of the small holes increases the likelihood of measuring transmission through the Fano-resonant Jaynes-Cummings system only.
\item A large hole defect at the center of the waveguide (blue) acts as the PTE~\cite{Yu2014-qp}, where tuning the radius to $r_\mathrm{PTE}=0.36d$ allows for an optimal destructive interference in the Fano resonance.
\item Several lines of shifts around the waveguide ($0.15d$ nearest neighbor [purple] and $0.1d$ second-nearest neighbor [green] away from the waveguide's center) tune its density of states to be centered around the cavity resonance and several hole shifts around the cavity (yellow)~\cite{Akahane2003-ma} optimize its intrinsic $Q_i\approx51,000$ (or rate $\kappa_i=\omega/Q_i$).
\item The total quality factor including the waveguide coupling is $Q_t\approx20,000$, such that the majority of the power injected into the cavity is transmitted to the output channel. Note, any quantum mechanically scattered light from a Jaynes-Cummings system emits equally into the input and output waveguides because unlike the injected classical light, it has no reflected partner with which to interferometrically cancel.
\end{itemize}
With these features in total, the system produces a Fano-resonant transmission plot (figure~\ref{figure:1}d) [with lattice constant $a=241\,\textrm{nm}$, as might be used in conjunction with traditional GaAs-based devices]. A well-tuned Fano resonance is critical for enabling a so-called self-homodyne interference. Specifically, when the Fano lineshape is tuned such that the dip approaches zero (as verified in a classical FDTD simulation), then all the incident classical light is interferometrically canceled. This then allows for the preferential emission of nonclassical light into the output channel (in a hypothetical experiment or quantum optical simulation)~\cite{Fischer2016-dj}.

% ------------------------------------------ section ------------------------------------------ 
%%%%%%%%%%%%%%%%%%%%%%%%%%%%%%%%%%%%

\section{Quantum optical modeling}

In this section, we additionally consider the effects of an embedded quantum emitter and discuss the ability of the Fano resonance to significantly modify the quantum statistics of the output light from the QD PCC system. First, we begin by discussing a quantum optical model that assumes the PTE acts to fully block the incident light, rather than transmitting some portion. This way we make comparisons to how many previous experiments operated~\cite{Faraon2008-zh,Reinhard2011-ye} and can introduce several concepts without the added complexity of the Fano interference. Then, we discuss how the tuned Fano resonances for optimal self-homodyne interference allow for preferential transmission of just the light with interesting quantum states.

% ---------------------------------------- sub-section ---------------------------------------- 
%%%%%%%%%%%%%%%%%%%%%%%%%%%%%%%%%%%%

\subsection{Pure Jaynes-Cummings emission}

When a quantum two-level system such as a neutrally-charged quantum dot couples to a cavity mode via a dipole coupling, the system is referred to as a Jaynes-Cummings system~\cite{Laussy2012-hr,Yoshie2004-yd,Englund2007-wy,Faraon2008-zh,Reinhard2011-ye}. Such a system is described by the Hamiltonian
\begin{equation}
H=\hbar\omega_a a^\dagger a + \hbar(\omega_a + \Delta)\sigma^\dagger \sigma + \hbar g (a^\dagger \sigma  + a \sigma^\dagger)\textrm{,}
\end{equation}
where $\omega_a$ denotes the frequency of the cavity mode, $a$ the annihilation operator associated with the cavity mode, $\sigma$ the lowering operator of the quantum two-level system, $\Delta$  the detuning between quantum emitter and cavity, and $g$ the emitter-cavity coupling strength. We have chosen to run our simulations with an easily achievable value~\cite{Yoshie2004-yd,Englund2007-wy,Faraon2008-zh} for the emitter-cavity coupling strength of $g=10\cdot 2\pi\,\textrm{GHz}$. Any realistic Jaynes-Cummings system also interacts with the outside world, both through the cavity energy decay (rate $\kappa$) and the quantum emitter decay (rate $\gamma$). Because our quantum emitter, an uncharged quantum dot, is embedded within a photonic bandgap, its coupling to free space via modes other than the discrete cavity channel is suppressed~\cite{Muller2015-il,Englund2005-wz} (i.e. we take~$\gamma\rightarrow 0$, see Appendix \ref{appendix:A} for a numerical justification).

\begin{figure}
  \includegraphics[width=\columnwidth]{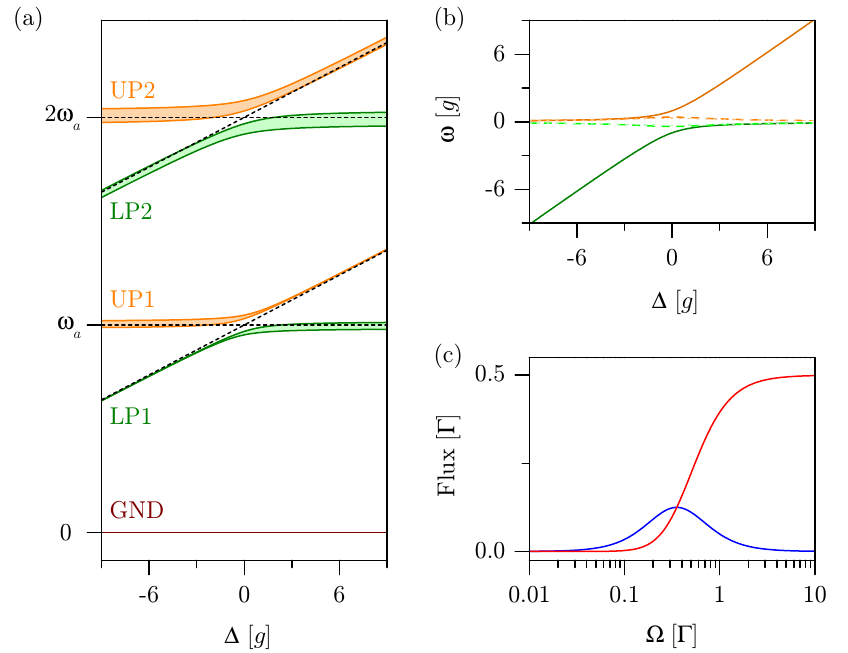}
  \caption{(a) Energies and linewidths of the polaritons (UP$n$ and LP$n$) of the first two rungs of the Jaynes-Cummings ladder (with emitter decay ${\gamma\rightarrow 0}$). Orange and green denote the upper and lower branches, respectively. Bounding regions depict the full-width half-maxes of the eigenstate linewidths. The ground state (GND) at zero energy is depicted as the brown line. (b) Transient energies to climb the ladder of Jaynes-Cummings eigenstates one-by-one, relative to the cavity's emission energy. Jumps up the upper and lower branches of the ladder are shown as orange and green, respectively. The first jumps are shown with the solid curves (from GND$\rightarrow$UP1 or GND$\rightarrow$LP1), second jumps with the dashed curves (from UP1$\rightarrow$UP2 or LP1$\rightarrow$LP2), and third jumps with the dotted curves (from UP2$\rightarrow$UP3 or LP2$\rightarrow$LP3). (c) Coherent (blue) and incoherent (red) portions of the emission from a quantum two-level system as a function of driving strength ($\propto E_\textrm{field}$), computed with the coherent drive on resonance with the transition.}
  \label{figure:2}
\end{figure}

If the cavity loss terms are included as damping to an effective Hamiltonian~\cite{Laussy2012-hr}, we can gain significant insight into how our system will behave simply by looking at its complex eigenvalues or polaritons; we show these eigenvalues schematically in figure~\ref{figure:2}a. Here, the orange and green colors indicate the upper (UP) and lower (LP) branches of the Jaynes-Cummings ladder, respectively, while the bounding curves that surround the eigenvalues depict the full-width half-maxes of their linewidths. Additionally, the ground state (GND) is depicted at zero energy by the brown line. The quantum emitter has a very strong effect on the loss rates in the first manifold for large $\Delta$ (i.e. LP1 and UP1), which can be seen in the rapid change of height in the orange and green bounding regions around $\omega_a$. However, the emitter's effect is much smaller for the second manifold (i.e. LP2 and UP2), where the orange and green bounding regions maintain relatively constant heights around $2\omega_a$. This behavior occurs because both detuned polaritons of the second manifold each contain an additional cavity photon that leaves the system at roughly a rate of~$\kappa$.

Another way to visualize this information is shown in figure~\ref{figure:2}b, where the transient energies to climb the Jaynes-Cummings ladder rung-by-rung~\cite{Laussy2012-hr,Radulaski2017-hj} are shown (either following GND$\rightarrow$UP1$\rightarrow$UP2$\rightarrow$UP3 or GND$\rightarrow$LP1$\rightarrow$LP2$\rightarrow$LP3	, which are the only transitions allowed since $\gamma\rightarrow0$). Here, the differential energies (relative to the cavity energy) that are required to add subsequent excitations to the ladder are shown, for both the upper and lower branches. For non-zero detunings, the branches are called emitter- or cavity-like if their linewidths trend towards $\gamma$ or $\kappa$, respectively. Note that especially for large detunings the energy levels in the cavity-like branch (LP$n$ for $\Delta>0$ and UP$n$ for $\Delta<0$, with $n>0$), and even those in the emitter-like branch above the first rung (UP$n$ for $\Delta>0$ and LP$n$ for $\Delta<0$, with $n>1$), are all approximately evenly spaced---\textit{we will refer to these levels as the harmonic portion of the Hamiltonian}. From these figures alone, we will explain the primary mechanism through which the Jaynes-Cummings ladder can act as a single-photon source, a phenomenon known as photon blockade~\cite{Muller2015-il,Faraon2008-zh,Reinhard2011-ye}.

In photon blockade the Jaynes-Cummings system is continuously driven by a coherent state flux $\ket{\alpha}$ resonant with the first polariton of the ladder, and its anharmonic energy levels filter the photon number of the pulse, transmitting light with a sub-Poisson photon number distribution. In the ideal case, only a single photon is transmitted by the system at a time---then the system effectively behaves like a two-level system. From the eigenenergy plots in figure~\ref{figure:2} one can see that for larger detunings, the anharmonicity of the Jaynes-Cummings ladder increases around the first emitter-like polariton. After absorption of the first photon by the system, a higher anharmonicity means that the energy to the next highest eigenstate able to absorb a photon is further away, and hence one would expect the single-photon purity to increase~\cite{Muller2015-il}. This increase in purity is quantified by a decrease in the second-order coherence statistic at zero delay, i.e.
\begin{equation}
g^{(2)}(0)=\frac{\langle a^\dagger a^\dagger a a \rangle}{\langle a^\dagger a \rangle^2}<1\textrm{,}
\end{equation}
which is referred to as anti-bunching. Although the degree of anti-bunching certainly suggests how well the Jaynes-Cummings system filters the incident pulse to ideal sub-Poissonian counting statistics, a strongly anti-bunched second-order coherence is not unique to emission from a two-level system~\cite{Zubizarreta_Casalengua2016-wj} nor sufficient to characterize single-photon emission~\cite{Camilo-xn}.

A more specific way to identify the behavior of a two-level system lies within the ratio of its coherently (classically) to incoherently (nonclassically) emitted light, as a function of driving strength of the coherent excitation~\cite{Loudon2000-li,Fischer2016-dj}. The coherent label is assigned to the light emission due to the square of a system operator's average transition dipole moment, e.g. $I_\textrm{c}\propto\langle a \rangle^2$, and the incoherent label is assigned to the light emission due to the average of the square of a system operator's dipole moment (known as an operator's quantum fluctuations), e.g. $I_\textrm{inc}\propto\langle a^2 \rangle-\langle a \rangle^2$. For the sake of comparison, we plot the portions of coherently and incoherently emitted light from a quantum two-level system in figure~\ref{figure:2}c. We will be re-plotting this set of values consistently in the next figure~(\ref{figure:3}) for comparison to how the Jaynes-Cummings system behaves. In doing so, we will quantify how closely the excitation of a polariton in a Jaynes-Cummings system matches that of a two-level system, i.e. how closely we truly excite only one polariton in the photon-blockade regime.

\begin{figure*}[!t]
  \includegraphics[width=\textwidth]{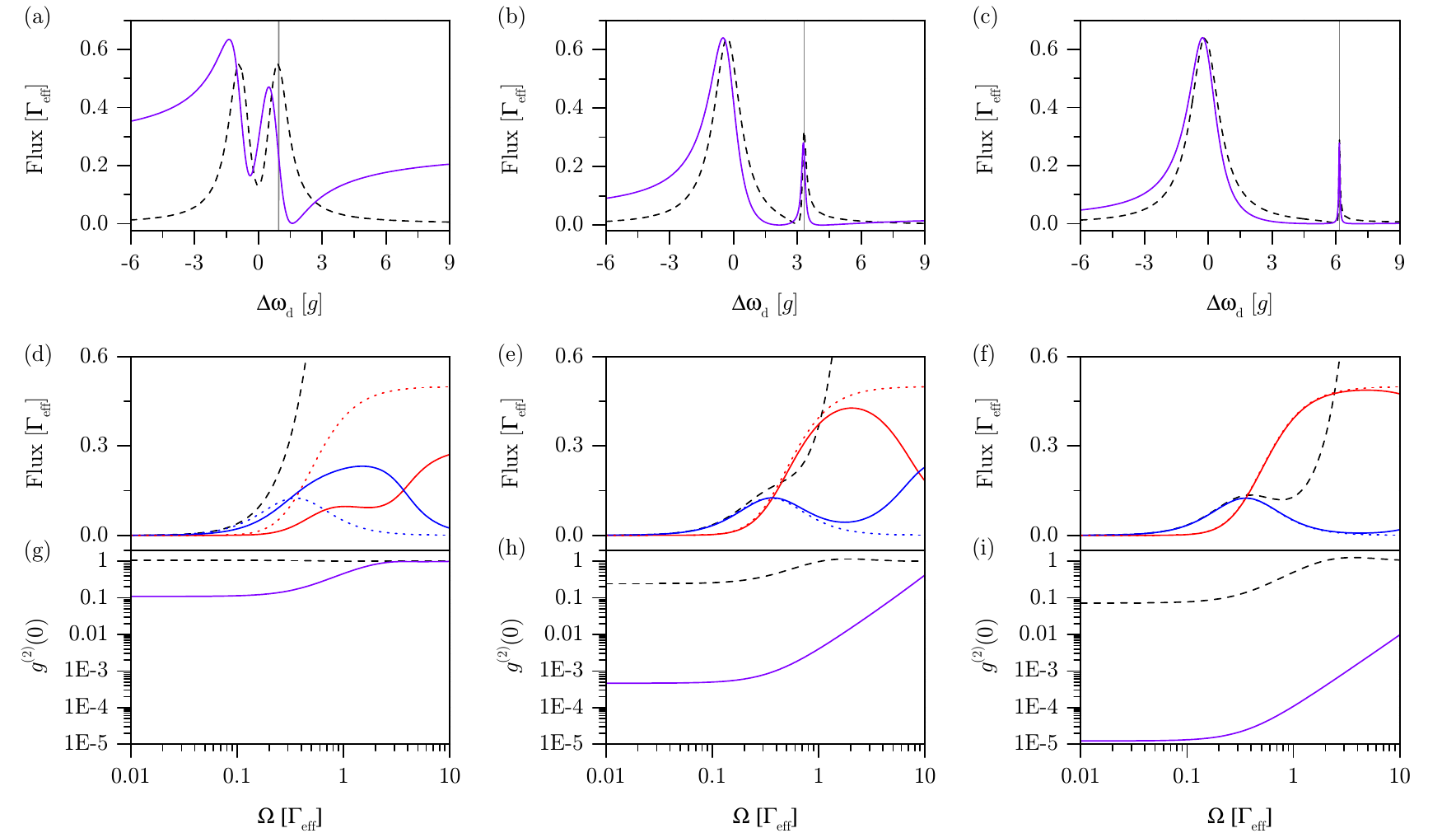}
  \caption{(a-c) Transmission through Jaynes-Cummings systems with detunings of (a) $\Delta=0$, (b) $\Delta=3g$, and (c) $\Delta=6g$. Dashed black curves represent the case where the PTE is fully blocking and only emission from the Jaynes-Cummings sytem is measured. Solid purple curves indicate Fano-resonant lineshapes with the interference optimally tuned for the emission of quantum light when the driving laser is resonant with the higher energy polariton, UP1 (denoted by solid vertical gray lines). (d-f) Decomposition of light from (a-c) into coherent and incoherent components when system is excited under the conditions of the vertical gray lines (a-c), respectively. Dashed black curves indicate the coherent emission from the Jaynes-Cummings system alone, solid red curves indicate incoherent emission from the Jaynes-Cummings system, while solid blue curves indicate the coherent portion of the transmitted light under optimal Fano-induced self-homodyne conditions in (a-c), respectively. Dashed blue and red curves indicate reference coherent and incoherent emission from a two-level system, respectively. (g-i) Second-order coherence statistics of total transmitted light when system is excited under the conditions of the vertical gray curves (a-c), respectively. Dashed black and solid purple curves indicate without and with the optimal homodyne interference, respectively. For (d-i), the driving strengths are normalized by the loss rates of the driven polariton.}
  \label{figure:3}
\end{figure*}

We now discuss the transmission through the Jaynes-Cummings system in the case where the PTE is fully blocking---we note this will only be a subset of the curves plotted in figure~\ref{figure:3}. In this case, all output emission is governed solely by the system operator $a$ (as is known from input-output theory~\cite{Gardiner2004-qu}). For the following analysis, we have chosen to analyze the transmitted light at three different emitter-cavity detunings: resonant $\Delta=0$, moderate $\Delta=3g$, and relatively large $\Delta=6g$. The transmission curves for asymptotically low driving strengths are shown as the dashed black curves in figures~\ref{figure:3}a-c, which are the standard Jaynes-Cummings transmission plots. As the detuning is increased, the linewidth of the emitter-like polariton decreases and trends toward the character of the quantum emitter. Then, we chose to resonantly excite the emitter-like polariton (UP1 since $\Delta>0$) in each case to study the quantum aspects of its emission, with the resonance conditions denoted by the vertical solid gray lines.

Thus in figures~\ref{figure:3}d-f, we decomposed the emission depicted in figures~\ref{figure:3}a-c into coherent and incoherent components when the system is excited under the conditions of the vertical gray lines in figures~\ref{figure:3}a-c, respectively, as a function of the system driving strengths. Importantly, we have normalized both the driving strengths and emission fluxes to the loss rates of the driven polaritons~\cite{Laussy2012-hr} where
\begin{equation}
\Gamma_\textrm{eff} = \frac{\kappa}{2}+2\,\textrm{Im} \left\{\sqrt{ g^2-\left( \frac{\kappa}{4}+\frac{\textbf{i}\Delta}{2} \right)^2 }\right\}\textrm{,}
\end{equation}
 so we can directly make comparisons to the decomposed emission from a two-level system. The dashed black and solid red curves indicate the coherent and incoherent emission from the Jaynes-Cummings systems, respectively. Meanwhile, the dashed blue and dashed red curves indicate reference coherent and incoherent emission from a two-level system, respectively. We can immediately note several trends:
\begin{itemize}
\item At arbitrarily low driving strengths, the coherent and incoherent emission proportions from the Jaynes-Cummings systems trend towards those of the two-level system for all detunings.
\item As the driving strengths increase, the emissions from all the studied systems deviate from the coherent and incoherent portions of emission from a two-level system. However, the larger the detuning, the slower the deviation from the behavior of a two-level system.
\item Nevertheless, even at the large detuning of $\Delta=6g$, for just moderate driving strengths of $\approx\Gamma_\textrm{eff}$ the coherently scattered light (dashed black line) begins to completely dominate. Thus, the driving powers for which the emission of any dissipative Jaynes-Cummings system matches that of a two-level system is extremely limited and prototypical quantum features such as saturation or Rabi oscillations should not be expected.
\end{itemize}

Additionally, we can look at the second-order coherence statistics to determine how well the systems are filtering the photon number distribution. Emission from an ideal two-level system perfectly filters the light such that $g^{(2)}(0)=0$, but the Jaynes-Cummings systems by themselves do a poor job at filtering the emission even for large detunings. The results of this filtering are shown in the dashed black curves of figures~\ref{figure:3}g-i, where the minimal values are $g^{(2)}(0)>0.08$. The imperfect performance in our simulation results may initially be surprising, since previous works proposed the Jaynes-Cummings polaritons as excellent two-level systems~\cite{Laussy2012-hr,Yoshie2004-yd,Englund2007-wy,Faraon2008-zh,Reinhard2011-ye}. Instead, many of these discussions have not fully taken the dissipative nature of experimental Jaynes-Cummings systems into account. Rather, the large linewidths of each of the rungs in the ladder (figure~\ref{figure:2}a) result in significant excitation to higher rungs of the ladder (which are harmonic and hence scatter coherently) even when the laser is only in direct resonance with the first polaritonic rung~\cite{Muller2015-il,Fischer2016-dj}. Next, we will fully justify how our proposed integrated QD PCC structure, inspired by previous works utilizing self-homodyne interference, can overcome these limitations in order to act as a nearly ideal two-level system.

% ---------------------------------------- sub-section ---------------------------------------- 
%%%%%%%%%%%%%%%%%%%%%%%%%%%%%%%%%%%%

\subsection{Emission including the Fano resonance}

Our previous modeling section considered the case where the PTE is fully blocking and only light scattered through the Jaynes-Cummings system is collected in the output waveguide. In this section, we relax this criterion and actively tune the PTE until the classically scattered light through the harmonic portion of the Jaynes-Cummings system is interferometrically canceled. Thus, we will detail a fully quantum-mechanical model that combines this Fano-resonant behavior with quantum optical modeling to numerically characterize the nonclassically emitted light. 

Before, the flux operator representing the emitted light into the waveguide was $\propto a$, however now it is~${\propto \sqrt{\kappa}/2\, a + t_B\alpha}$, where $t_B$ is the transmission coefficient of the PTE and $\alpha$ again represents the incident coherent flux. Defining the new operator $b=\sqrt{\kappa}/2\,a + t_B\alpha$ to represent this interference, we again re-evaluate our previous expressions. While before the steady-state transmitted flux was given by $T=\kappa/2\langle a^\dagger a \rangle$, it is now given by $T=\langle b^\dagger b \rangle$. By a similar operator replacement, the second-order coherence statistic is now given by
\begin{equation}
g^{(2)}(0)=\frac{\langle b^\dagger b^\dagger b b \rangle}{\langle b^\dagger b \rangle^2}\textrm{.}
\end{equation}
The interference present in the operator $b$ now allows for the alteration of the measured portion of the coherently scattered light, though it leaves the incoherent portion unchanged since the incident flux has only a coherent portion. Using these new waveguide operators, we calculated the optimal values for the $t_B$'s by passing the incident flux $\ket{\alpha}$ through a reference cavity with no light-matter interaction; then $t_B\alpha=-\sqrt{\kappa}/2\langle a_\textrm{ref} \rangle$.

Therefore, our proposed device performs a homodyne measurement~\cite{Carmichael2009-nj} of the actual mode operator~$a$. However, unlike previously suggested~\cite{Fischer2016-dj} it does not perfectly isolate the incoherently scattered portion of light from the Jaynes-Cummings system. Were this the case, the emitted light would not actually anti-bunch and certainly would not behave like it were emitted from a two-level system. \textit{Instead, the self-homodyne interference only removes the coherent light scattered by the harmonic portion of the Jaynes-Cummings system. This way the light coherently scattered by the first emitter-like polariton is left untouched, which is crucial for anti-bunching}. The reason it is left untouched when this interference is optimized (as above) is that the coherently scattered light from the emitter-like polariton has a dramatically different phase than that of the cavity-scattered coherent light. Here the cavity-like branches are far off resonance, so they scatter with little phase shift, while the emitter-like polariton scatters with a phase shift of $\pi/2$ since it is on resonance. Because at high powers the cavity-scattered light completely dominates, it is unsurprising that without fully understanding this subtle point, previous experiments with self-homodyne interference were still able to improve the quality of the single-photon emission. Armed with this information, we can now correctly analyze how self-homodyne interference can change the emission character of a CQED system as compared to that from a two-level system. 

We now revisit figure~\ref{figure:3} and discuss the results of the simulations that include the Fano and self-homodyne interference. First consider the transmission plots through the Jaynes-Cummings system with the optimally determined PTEs (figures~\ref{figure:3}a-c): strong asymmetries are now present due to the direct transmission channels in each case (purple curves). The Fano-resonant Jaynes-Cummings transmission plots show decreased amplitudes directly on the polaritons indicated by the vertical gray lines, with the largest percentage change in the case of zero detuning (figure~\ref{figure:3}a) since the portion of harmonic scattering is largest here. However, the changes are even more interesting when considering the coherently scattered portion as a function of the driving strength (solid blue curves in figures~\ref{figure:3}d-f). Now, the coherently scattered portions better match those from a two-level system of equivalent loss rate. For zero detuning, there is still significant disparity at high drive powers but low powers match fairly well (figure~\ref{figure:3}a). Notably for the large detuning case, the emission is now nearly identical to that from a two-level system even for extremely strong driving strengths (up to $10\Gamma_\textrm{eff}$).

Similarly dramatic are the changes in the second-order coherence statistics for the transmission through the Fano-resonant Jaynes-Cummings systems. Just as in their coherently scattered ratios, the second-order coherence statistics now markedly improve towards the ideal single-photon emission regime of $g^{(2)}(0)=0$ (compare figures~\ref{figure:3}g-i where the dashed black and solid purple curves indicate without and with the optimal Fano-induced self-homodyne interference, respectively). In all cases this improvement with interference in transmission is at least an order of magnitude better. Even though the large detuning case might have initially seemed like it would not improve much in $g^{(2)}(0)$, we find it very notable that removing any unwanted coherent scattering by the Jaynes-Cummings system itself can improve the single-photon error rate by almost four orders of magnitude. Clearly, the self-homodyne interference can play a powerful role in ensuring that the emission from the emitter-like polariton in a Jaynes-Cummings system actually behaves like emission from an ideal two-level system.

We also now consider an interesting feature in the second-order coherence curves that we believe to be elucidating. Again consider figures~\ref{figure:3}d-f, but note that the $g^{(2)}(0)$ curves plateau at low driving strengths and begin to rise at exactly the same normalized driving strengths. Notably, this rise occurs precisely when the emitter-like polariton begins to saturate, i.e. when $I_\textrm{inc}>I_\textrm{c}$. From this observation, we can conclude that the minima of $g^{(2)}(0)$ are not due to excitation of polaritons above the emitter-like polaritons. Instead, the minima are determined by the slight anharmonicity of the cavity-like polariton branches, whose emission cannot be fully canceled by a coherent beam.

\begin{figure}
  \includegraphics[width=\columnwidth]{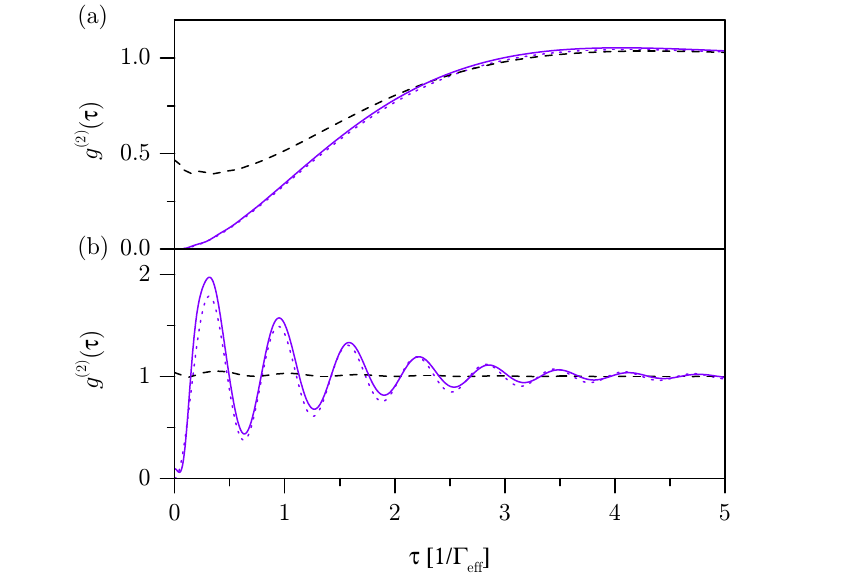}
  \caption{Second-order coherence as a function of time delay at moderate detuning ($\Delta=3g$) with moderate [$0.4\Gamma_\textrm{eff}$] (a) and relatively large [$5\Gamma_\textrm{eff}$] (b) driving strengths. Dashed black curves represent cases for pure Jaynes-Cummings emission, solid purple curves include self-homodyne interference, and dotted purpled curves depict ideal two-level system behavior.}
  \label{figure:4}
\end{figure}

Finally, we briefly consider the dynamical issue of Rabi oscillations. Although we only consider steady-state phenomenon in this work, we can look at steady-state quantum correlations that are closely tied to Rabi oscillations. This signature is related to the Mollow triplet and appears in the oscillations of the second-order coherence as a function of time delay~\cite{Carmichael2009-nj}, i.e. in
\begin{equation}
g^{(2)}(\tau)=\lim_{t\rightarrow\infty}\frac{\langle b^\dagger(t)b^\dagger(t+\tau)b(t+\tau)b(t)\rangle}{\langle b^\dagger(t)b(t)\rangle^2}\textrm{,}
\end{equation}
but with $b(t)=\sqrt{\kappa}/2\left(a(t) - \langle a_\textrm{ref} \rangle\right)$ for the self-homodyne interference since the input coherent flux has uncorrelated statistics.
We now consider the delayed second-order coherences of the emission from our Jaynes-Cummings system at moderate detuning (dashed black and solid purple curves in figure~\ref{figure:4}) and compare them to the second-order coherences of emission from an ideal two-level system (dotted purple curves in figure~\ref{figure:4}). For both moderate and high powers at the moderate detuning of $\Delta=3g$, the self-homodyne interference very clearly allows the two-level nature of the Jaynes-Cummings emission to be revealed (solid purple curves) from what otherwise would be almost trivial counting statistics, i.e. $g^{(2)}(\tau)\approx1$ (dashed black lines). In our previous work where we investigated dissipative Jaynes-Cummings systems based on QDs strongly coupled to PCCs~\cite{Muller2015-om,Dory2016-rf}, we observed Rabi oscillations which require high driving strengths. As the current work and results of figure~\ref{figure:4}b suggest, it is thanks to the self-homodyne interference that these oscillations were observable.

% ------------------------------------------- section ------------------------------------------ 
%%%%%%%%%%%%%%%%%%%%%%%%%%%%%%%%%%%%

\subsection{Multi-photon resonances}

\begin{figure}
  \includegraphics[width=\columnwidth]{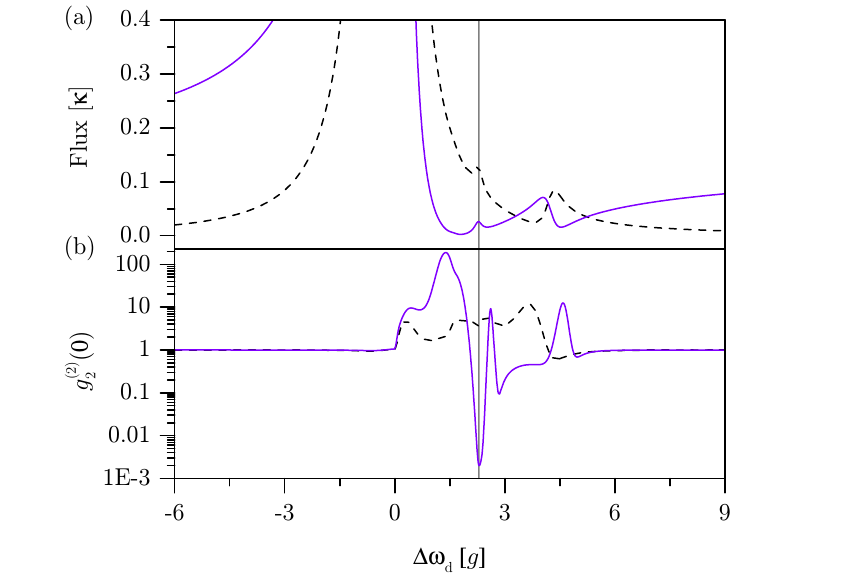}
  \caption{Detuned Jaynes-Cummings system under strong drive ($\Delta=4g$), with (a) transmission and the (b) two-photon bundling statistic that highlights presence of a two-photon resonance. Colors indicate without (dashed black line) and with (purple line) optimally tuned self-homodyne interference. Vertical gray line indicates frequency of two-photon resonance in transmission.}
  \label{figure:5}
\end{figure}

Up until this point, we have considered the self-homodyne interference as a method for optimizing the quantum two-level type behavior of the emitter-like polariton in a Jaynes-Cummings system. In principle, the resulting single-photon emission could be generated either with a weakly coupled CQED system or simply a two-level system, after properly rejecting the coherent driving light. For this section, we explore phenomena unique to the complex ladder of Jaynes-Cummings levels: its multi-photon resonances. While these resonances have been explored in super-conducting systems~\cite{Schuster2008-go}, they have yet to be explored in solid-state systems at optical frequencies. Here, we show how self-homodyne interference can lead to the direct observation of two-photon resonances in our proposed integrated platform, and comment on its potential for observation of higher-order resonances as well.

Consider our on-chip Jaynes-Cummings system when moderately detuned ($\Delta=4g$), but with a state-of-the-art $g/\kappa=2.4$ ratio~\cite{Arakawa2012-yd}. Unlike for a moderate $g/\kappa$ system, as the driving field strength is increased beyond saturation of the emitter-like polariton, a two-photon resonance begins to appear at $2.29g$. This resonance occurs between GND$\leftrightarrow$UP2 with UP1 as an off-resonant intermediate state, and at the frequency that splits the difference between the transition energies of GND$\leftrightarrow$UP1 and UP1$\leftrightarrow$UP2.

However, its transmission spectrum (dashed black line in figure~\ref{figure:5}a) only barely reveals the two-photon resonance (denoted by the gray line) due to an excess of coherent scattering from the harmonic portion of the Hamiltonian. Fortunately, the optimally tuned self-homodyne interference mostly mitigates this unwanted emission making the resonance easier to observe (solid purple line). Even more stark is the contrast with and without the interference to the second-order photon \textit{bundling} statistics (figure~\ref{figure:5}b), of the form~\cite{Sanchez_Munoz2014-zk}
\begin{equation}
g^{(2)}_2(0) = \frac{\langle (a^\dagger)^4 (a)^4 \rangle}{\langle (a^\dagger)^2 (a)^2 \rangle^2}\textrm{.}
\end{equation}
The second-order bundling statistic, unlike the standard $n$-th order coherence statistics, is a valuable indicator of the presence of two-photon resonances in the Jaynes-Cummings ladder when it shows anti-bunching. Specifically, $g^{(2)}_n(0)$ looks at how bundles of $n$ photons anti-bunch, rather than just anti-bunching of single photons. Now the value of the self-homodyne interference becomes exceedingly clear by allowing for strong anti-bunching in $g^{(2)}_2(0)$ [solid purple line], which would not even be observed without the interference (dashed black line). Further enhancements in the $g/\kappa$ ratio, which may be enabled by new developments in GaAs surface engineering~\cite{Guha2017-qy}, will enable observation of multi-photon resonances involving more than two photons.

Finally, we briefly discuss the output states resulting from the multi-photon resonances under self-homodyne measurement. These resonances actually result in a somewhat complicated emission, which we believe would be an interesting subject for future research. The main observation is that even though the photons are absorbed at the same frequency, the detuned Jaynes-Cummings ladder emits them at different frequencies. Take the two-photon resonance: The energy leaves UP2 via a cascade where the first photon is emitted at the cavity-like polariton's frequency (UP2$\rightarrow$UP1 transition) and the second photon is emitted at the emitter-like polariton's frequency (UP1$\rightarrow$GND transition). Higher $n$-photon resonances emit roughly $n$-1 photons at the cavity-like frequency followed by a single photon at the emitter-like frequency. This situation is reminiscent of the photon bundling in the work of S\'{a}nchez Mu\~{n}oz et al. \cite{Sanchez_Munoz2014-zk}, but interestingly occurs under weak driving and at lower detunings. Since the light is emitted at multiple frequencies through the same output channel, a more complicated analysis that includes frequency filtering will be required to quantitatively study the output emission.

% ------------------------------------------- section ------------------------------------------ 
%%%%%%%%%%%%%%%%%%%%%%%%%%%%%%%%%%%%

\subsection{Conclusions}
Thus, we have both shown and illuminated how self-homodyne interference can lead to a plethora of interesting CQED effects. We showed how the self-homodyne interference generated from an optimally tuned Fano resonance can be utilized to strongly enhance the single-photon emission purity of a dissipative CQED system and allow for the observation of Rabi oscillations between the ground state and the emitter-like polariton. We have previously verified some of these effects with off-chip configurations utilizing self-homodyne interference~\cite{Fischer2016-dj,Muller2016-fs,Dory2016-rf}, but here we have thoroughly explained the enhancement in counting statistics and in a fully integrated on-chip geometry. Furthermore, we showed how these effects are completely compatible with an integrated waveguide structure and should soon be within sight of an experimental demonstration. Although QD-phonon interaction has recently been shown to be a limiting factor in the indistinguishability of single-photon emission \cite{Iles-Smith2016-bp,Grange2016-yp}, for strongly-coupled systems the dephasing results in energy transfer to the cavity-like polariton, which could potentially be completely removed by a photonic crystal drop filter to circumvent this limit \cite{Muller2016-fs}. Based on our proposed device's ease of on-chip integratability and on the power of an optimally tuned Fano interference to bring orders of magnitude of improvement to already well-performing CQED systems, we expect this type of quantum interference to readily find its way into photonic integrated circuits in the near future.

% ------------------------------------------- section ------------------------------------------ 
%%%%%%%%%%%%%%%%%%%%%%%%%%%%%%%%%%%%

\subsection{Acknowledgments}

We gratefully acknowledge financial support from the National Science Foundation (Division of Materials Research - Grant. No. 1503759) and the Bavaria California Technology Center (BaCaTeC). KAF acknowledges support from the Lu Stanford Graduate Fellowship and the National Defense Science and Engineering Graduate Fellowship. YK acknowledges support from the Art and Mary Fong Stanford Graduate Fellowship and the National Defense Science and Engineering Graduate Fellowship. CD acknowledges support from the Andreas Bechtolsheim Stanford Graduate Fellowship. KM acknowledges support from the Alexander von Humboldt Foundation.

% ------------------------------------------- appendices -------------------------------------- 
%%%%%%%%%%%%%%%%%%%%%%%%%%%%%%%%%%%%

\appendix

\section{Coupling of embedded dipole to leaky modes}\label{appendix:A}

\begin{figure}[]
  \includegraphics[width=\columnwidth]{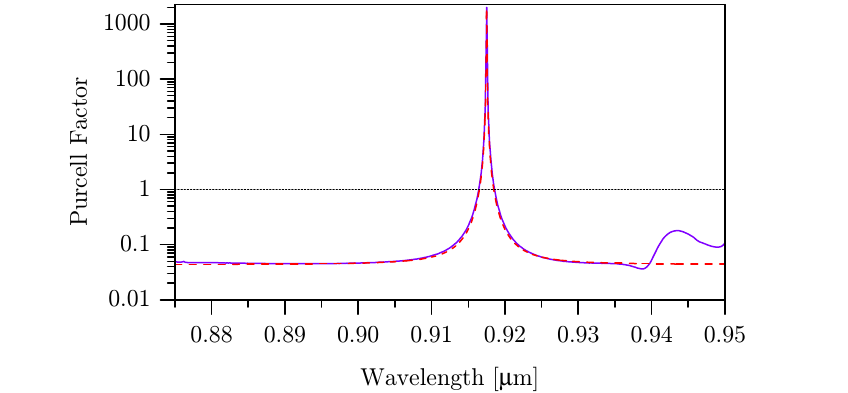}
  \caption{Purcell factor for a dipole located at the center of the cavity in our proposed self-homodyne architecture (purple curve). The Purcell factor is fitted well with a Lorentzian plus a constant (red dashed curve). Purcell suppression corresponds to values below the black dashed line.}
  \label{figure:A1}
\end{figure}

In the main text, we argued that an uncharged quantum dot is embedded within a photonic bandgap, and hence its coupling to free space via modes other than the discrete cavity channel is suppressed (i.e. we take~$\gamma\rightarrow 0$). Here, we show this approximation is valid by looking at the Purcell factor of a dipole embedded at the field maximum of the photonic crystal self-homodyne architecture (figure~\ref{figure:A1}). Because the Purcell factor (purple) is proportional to the local density of states (LDOS) over the bulk LDOS, then we can use it to determine the relative contributions to decay via the cavity compared to other leaky modes. From a quantum-mechanical perspective, the cavity channel is modeled by a Lorentzian LDOS while the other leaky modes are modeled by a flat LDOS. By fitting the Purcell factor to a constant plus a Lorentzian, we can quantify the relative contributions of each. Specifically, far away from the resonance the Purcell factor approaches a constant value of 0.043, which corresponds to Purcell suppression of emission through leaky modes by a factor of 23. While this factor is slightly lower than suppression by a bulk photonic crystal, it is consistent with calculated and measured suppression of a dipole embedded in an L3 photonic crystal cavity \cite{Englund2005-wz}.

\section{Minima of second-order coherence}

\begin{figure}[]
  \includegraphics[width=\columnwidth]{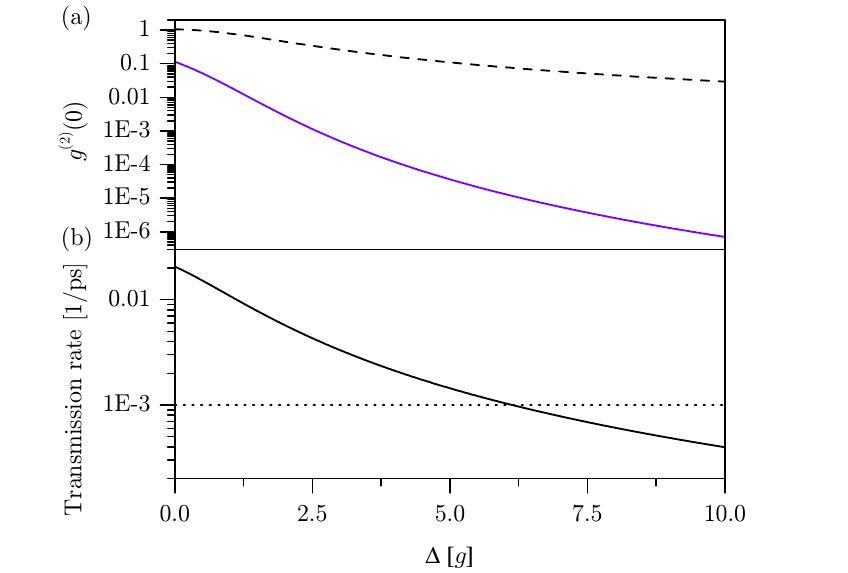}
  \caption{Detuning dependent single-photon generation under weak excitation. (a) Minimum values of $g^{(2)}(0)$ under weak excitation as a function of the detuning $\Delta$. Dashed black curve represents the case when exciting the emitter-like Polariton of a pure Jaynes-Cummings system, and the purple curve represents the same excitation conditions but with self-homodyne interference. (b) Transmission rate when transmitting through the emitter-like polariton (solid) of the strongly-coupled system. Values below dotted line are Purcell suppressed.\vspace{0pt}}
  \label{figure:A2}
\end{figure}

In the main text, we showed that the single-photon purity improves with increasing detuning (from left to right in figure 3). We now show that this trend is maintained for even larger detunings in figure~\ref{figure:A2}a, where the decreasing $g^{(2)}(0)$ and hence increasing purity with detuning is monotonic both with and without self-homodyne interference (under weak excitation). We also note that there is a trade-off between transmission rate and single-photon purity at large detunings. This can be seen in figure~\ref{figure:A2}b, where the transmission rate monotonically decreases with increasing detuning. For the parameters studied in the self-homodyne architecture, at a detuning of around $6.2g$ the emission becomes Purcell suppressed for a typical quantum dot with bulk lifetime 1\,ns.

% ------------------------------------------- bibliography ------------------------------------ 
%%%%%%%%%%%%%%%%%%%%%%%%%%%%%%%%%%%%

\bibliographystyle{apsrev4-1-PRX}
\bibliography{bibliography}% Produces the bibliography via BibTeX.

\begin{thebibliography}{48}%
\makeatletter
\providecommand \@ifxundefined [1]{%
 \@ifx{#1\undefined}
}%
\providecommand \@ifnum [1]{%
 \ifnum #1\expandafter \@firstoftwo
 \else \expandafter \@secondoftwo
 \fi
}%
\providecommand \@ifx [1]{%
 \ifx #1\expandafter \@firstoftwo
 \else \expandafter \@secondoftwo
 \fi
}%
\providecommand \natexlab [1]{#1}%
\providecommand \enquote  [1]{``#1''}%
\providecommand \bibnamefont  [1]{#1}%
\providecommand \bibfnamefont [1]{#1}%
\providecommand \citenamefont [1]{#1}%
\providecommand \href@noop [0]{\@secondoftwo}%
\providecommand \href [0]{\begingroup \@sanitize@url \@href}%
\providecommand \@href[1]{\@@startlink{#1}\@@href}%
\providecommand \@@href[1]{\endgroup#1\@@endlink}%
\providecommand \@sanitize@url [0]{\catcode `\\12\catcode `\$12\catcode
  `\&12\catcode `\#12\catcode `\^12\catcode `\_12\catcode `\%12\relax}%
\providecommand \@@startlink[1]{}%
\providecommand \@@endlink[0]{}%
\providecommand \url  [0]{\begingroup\@sanitize@url \@url }%
\providecommand \@url [1]{\endgroup\@href {#1}{\urlprefix }}%
\providecommand \urlprefix  [0]{URL }%
\providecommand \Eprint [0]{\href }%
\providecommand \doibase [0]{http://dx.doi.org/}%
\providecommand \selectlanguage [0]{\@gobble}%
\providecommand \bibinfo  [0]{\@secondoftwo}%
\providecommand \bibfield  [0]{\@secondoftwo}%
\providecommand \translation [1]{[#1]}%
\providecommand \BibitemOpen [0]{}%
\providecommand \bibitemStop [0]{}%
\providecommand \bibitemNoStop [0]{.\EOS\space}%
\providecommand \EOS [0]{\spacefactor3000\relax}%
\providecommand \BibitemShut  [1]{\csname bibitem#1\endcsname}%
\let\auto@bib@innerbib\@empty
%</preamble>
\bibitem [{\citenamefont {O'Brien}\ \emph {et~al.}(2009)\citenamefont
  {O'Brien}, \citenamefont {Furusawa},\ and\ \citenamefont
  {Vu\v{c}kovi\'{c}}}]{OBrien2009-dq}%
  \BibitemOpen
  \bibfield  {author} {\bibinfo {author} {\bibfnamefont {J.~L.}\ \bibnamefont
  {O'Brien}}, \bibinfo {author} {\bibfnamefont {A.}~\bibnamefont {Furusawa}}, \
  and\ \bibinfo {author} {\bibfnamefont {J.}~\bibnamefont {Vu\v{c}kovi\'{c}}},\
  }\bibfield  {title} {\emph {\bibinfo {title} {Photonic quantum
  technologies},\ }}\href@noop {} {\bibfield  {journal} {\bibinfo  {journal}
  {Nat. Photonics}\ }\textbf {\bibinfo {volume} {3}},\ \bibinfo {pages} {687}
  (\bibinfo {year} {2009})}\BibitemShut {NoStop}%
\bibitem [{\citenamefont {Soref}(2006)}]{Soref2006-ae}%
  \BibitemOpen
  \bibfield  {author} {\bibinfo {author} {\bibfnamefont {R.}~\bibnamefont
  {Soref}},\ }\bibfield  {title} {\emph {\bibinfo {title} {The past, present,
  and future of silicon photonics},\ }}\href@noop {} {\bibfield  {journal}
  {\bibinfo  {journal} {IEEE J. Sel. Top. Quantum Electron.}\ }\textbf
  {\bibinfo {volume} {12}},\ \bibinfo {pages} {1678} (\bibinfo {year}
  {2006})}\BibitemShut {NoStop}%
\bibitem [{\citenamefont {Collins}\ \emph {et~al.}(2013)\citenamefont
  {Collins}, \citenamefont {Xiong}, \citenamefont {Rey}, \citenamefont {Vo},
  \citenamefont {He}, \citenamefont {Shahnia}, \citenamefont {Reardon},
  \citenamefont {Krauss}, \citenamefont {Steel}, \citenamefont {Clark},\ and\
  \citenamefont {Eggleton}}]{Collins2013-qc}%
  \BibitemOpen
  \bibfield  {author} {\bibinfo {author} {\bibfnamefont {M.~J.}\ \bibnamefont
  {Collins}}, \bibinfo {author} {\bibfnamefont {C.}~\bibnamefont {Xiong}},
  \bibinfo {author} {\bibfnamefont {I.~H.}\ \bibnamefont {Rey}}, \bibinfo
  {author} {\bibfnamefont {T.~D.}\ \bibnamefont {Vo}}, \bibinfo {author}
  {\bibfnamefont {J.}~\bibnamefont {He}}, \bibinfo {author} {\bibfnamefont
  {S.}~\bibnamefont {Shahnia}}, \bibinfo {author} {\bibfnamefont
  {C.}~\bibnamefont {Reardon}}, \bibinfo {author} {\bibfnamefont {T.~F.}\
  \bibnamefont {Krauss}}, \bibinfo {author} {\bibfnamefont {M.~J.}\
  \bibnamefont {Steel}}, \bibinfo {author} {\bibfnamefont {A.~S.}\ \bibnamefont
  {Clark}}, \ and\ \bibinfo {author} {\bibfnamefont {B.~J.}\ \bibnamefont
  {Eggleton}},\ }\bibfield  {title} {\emph {\bibinfo {title} {Integrated
  spatial multiplexing of heralded single-photon sources},\ }}\href {\doibase
  10.1038/ncomms3582} {\bibfield  {journal} {\bibinfo  {journal} {Nat.
  Commun.}\ }\textbf {\bibinfo {volume} {4}},\ \bibinfo {pages} {2582}
  (\bibinfo {year} {2013})}\BibitemShut {NoStop}%
\bibitem [{\citenamefont {Galli}\ \emph {et~al.}(2014)\citenamefont {Galli},
  \citenamefont {Baehr-Jones}, \citenamefont {Hochberg}, \citenamefont
  {Englund},\ and\ \citenamefont {{others}}}]{Galli2014-bs}%
  \BibitemOpen
  \bibfield  {author} {\bibinfo {author} {\bibfnamefont {M.}~\bibnamefont
  {Galli}}, \bibinfo {author} {\bibfnamefont {T.}~\bibnamefont {Baehr-Jones}},
  \bibinfo {author} {\bibfnamefont {M.}~\bibnamefont {Hochberg}}, \bibinfo
  {author} {\bibfnamefont {D.}~\bibnamefont {Englund}}, \ and\ \bibinfo
  {author} {\bibnamefont {{others}}},\ }\bibfield  {title} {\emph {\bibinfo
  {title} {Integrated source of spectrally filtered correlated photons for
  large-scale quantum photonic systems},\ }}\href@noop {} {\bibfield  {journal}
  {\bibinfo  {journal} {Physical Review X}\ }\textbf {\bibinfo {volume} {4}},\
  \bibinfo {pages} {041047} (\bibinfo {year} {2014})}\BibitemShut {NoStop}%
\bibitem [{\citenamefont {Meany}\ \emph {et~al.}(2014)\citenamefont {Meany},
  \citenamefont {Ngah}, \citenamefont {Collins}, \citenamefont {Clark},
  \citenamefont {Williams}, \citenamefont {Eggleton}, \citenamefont {Steel},
  \citenamefont {Withford}, \citenamefont {Alibart},\ and\ \citenamefont
  {Tanzilli}}]{Meany2014-uu}%
  \BibitemOpen
  \bibfield  {author} {\bibinfo {author} {\bibfnamefont {T.}~\bibnamefont
  {Meany}}, \bibinfo {author} {\bibfnamefont {L.~A.}\ \bibnamefont {Ngah}},
  \bibinfo {author} {\bibfnamefont {M.~J.}\ \bibnamefont {Collins}}, \bibinfo
  {author} {\bibfnamefont {A.~S.}\ \bibnamefont {Clark}}, \bibinfo {author}
  {\bibfnamefont {R.~J.}\ \bibnamefont {Williams}}, \bibinfo {author}
  {\bibfnamefont {B.~J.}\ \bibnamefont {Eggleton}}, \bibinfo {author}
  {\bibfnamefont {M.~J.}\ \bibnamefont {Steel}}, \bibinfo {author}
  {\bibfnamefont {M.~J.}\ \bibnamefont {Withford}}, \bibinfo {author}
  {\bibfnamefont {O.}~\bibnamefont {Alibart}}, \ and\ \bibinfo {author}
  {\bibfnamefont {S.}~\bibnamefont {Tanzilli}},\ }\bibfield  {title} {\emph
  {\bibinfo {title} {Hybrid photonic circuit for multiplexed heralded single
  photons},\ }}\href@noop {} {\bibfield  {journal} {\bibinfo  {journal} {Laser
  Photonics Rev.}\ }\textbf {\bibinfo {volume} {8}},\ \bibinfo {pages} {L42}
  (\bibinfo {year} {2014})}\BibitemShut {NoStop}%
\bibitem [{\citenamefont {Aharonovich}\ \emph {et~al.}(2016)\citenamefont
  {Aharonovich}, \citenamefont {Englund},\ and\ \citenamefont
  {Toth}}]{Aharonovich2016-pq}%
  \BibitemOpen
  \bibfield  {author} {\bibinfo {author} {\bibfnamefont {I.}~\bibnamefont
  {Aharonovich}}, \bibinfo {author} {\bibfnamefont {D.}~\bibnamefont
  {Englund}}, \ and\ \bibinfo {author} {\bibfnamefont {M.}~\bibnamefont
  {Toth}},\ }\bibfield  {title} {\emph {\bibinfo {title} {Solid-state
  single-photon emitters},\ }}\href@noop {} {\bibfield  {journal} {\bibinfo
  {journal} {Nat. Photonics}\ }\textbf {\bibinfo {volume} {10}},\ \bibinfo
  {pages} {631} (\bibinfo {year} {2016})}\BibitemShut {NoStop}%
\bibitem [{\citenamefont {Faraon}\ \emph {et~al.}(2011)\citenamefont {Faraon},
  \citenamefont {Majumdar}, \citenamefont {Englund}, \citenamefont {Kim},
  \citenamefont {Bajcsy},\ and\ \citenamefont
  {Vu\v{c}kovi\'{c}}}]{Englund2011-lx}%
  \BibitemOpen
  \bibfield  {author} {\bibinfo {author} {\bibfnamefont {A.}~\bibnamefont
  {Faraon}}, \bibinfo {author} {\bibfnamefont {A.}~\bibnamefont {Majumdar}},
  \bibinfo {author} {\bibfnamefont {D.}~\bibnamefont {Englund}}, \bibinfo
  {author} {\bibfnamefont {E.}~\bibnamefont {Kim}}, \bibinfo {author}
  {\bibfnamefont {M.}~\bibnamefont {Bajcsy}}, \ and\ \bibinfo {author}
  {\bibfnamefont {J.}~\bibnamefont {Vu\v{c}kovi\'{c}}},\ }\bibfield  {title}
  {\emph {\bibinfo {title} {Integrated quantum optical networks based on
  quantum dots and photonic crystals},\ }}\href
  {http://stacks.iop.org/1367-2630/13/i=5/a=055025} {\bibfield  {journal}
  {\bibinfo  {journal} {New J. Phys.}\ }\textbf {\bibinfo {volume} {13}},\
  \bibinfo {pages} {055025} (\bibinfo {year} {2011})}\BibitemShut {NoStop}%
\bibitem [{\citenamefont {Arcari}\ \emph {et~al.}(2014)\citenamefont {Arcari},
  \citenamefont {S{\"{o}}llner}, \citenamefont {Javadi}, \citenamefont
  {Lindskov~Hansen}, \citenamefont {Mahmoodian}, \citenamefont {Liu},
  \citenamefont {Thyrrestrup}, \citenamefont {Lee}, \citenamefont {Song},
  \citenamefont {Stobbe},\ and\ \citenamefont {Lodahl}}]{Arcari2014-ch}%
  \BibitemOpen
  \bibfield  {author} {\bibinfo {author} {\bibfnamefont {M.}~\bibnamefont
  {Arcari}}, \bibinfo {author} {\bibfnamefont {I.}~\bibnamefont
  {S{\"{o}}llner}}, \bibinfo {author} {\bibfnamefont {A.}~\bibnamefont
  {Javadi}}, \bibinfo {author} {\bibfnamefont {S.}~\bibnamefont
  {Lindskov~Hansen}}, \bibinfo {author} {\bibfnamefont {S.}~\bibnamefont
  {Mahmoodian}}, \bibinfo {author} {\bibfnamefont {J.}~\bibnamefont {Liu}},
  \bibinfo {author} {\bibfnamefont {H.}~\bibnamefont {Thyrrestrup}}, \bibinfo
  {author} {\bibfnamefont {E.~H.}\ \bibnamefont {Lee}}, \bibinfo {author}
  {\bibfnamefont {J.~D.}\ \bibnamefont {Song}}, \bibinfo {author}
  {\bibfnamefont {S.}~\bibnamefont {Stobbe}}, \ and\ \bibinfo {author}
  {\bibfnamefont {P.}~\bibnamefont {Lodahl}},\ }\bibfield  {title} {\emph
  {\bibinfo {title} {Near-unity coupling efficiency of a quantum emitter to a
  photonic crystal waveguide},\ }}\href@noop {} {\bibfield  {journal} {\bibinfo
   {journal} {Phys. Rev. Lett.}\ }\textbf {\bibinfo {volume} {113}},\ \bibinfo
  {pages} {093603} (\bibinfo {year} {2014})}\BibitemShut {NoStop}%
\bibitem [{\citenamefont {Dietrich}\ \emph {et~al.}(2016)\citenamefont
  {Dietrich}, \citenamefont {Fiore}, \citenamefont {Thompson}, \citenamefont
  {Kamp},\ and\ \citenamefont {H\"{o}fling}}]{LPOR201500321}%
  \BibitemOpen
  \bibfield  {author} {\bibinfo {author} {\bibfnamefont {C.~P.}\ \bibnamefont
  {Dietrich}}, \bibinfo {author} {\bibfnamefont {A.}~\bibnamefont {Fiore}},
  \bibinfo {author} {\bibfnamefont {M.~G.}\ \bibnamefont {Thompson}}, \bibinfo
  {author} {\bibfnamefont {M.}~\bibnamefont {Kamp}}, \ and\ \bibinfo {author}
  {\bibfnamefont {S.}~\bibnamefont {H\"{o}fling}},\ }\bibfield  {title} {\emph
  {\bibinfo {title} {Ga{A}s integrated quantum photonics: Towards compact and
  multi-functional quantum photonic integrated circuits},\ }}\href {\doibase
  10.1002/lpor.201500321} {\bibfield  {journal} {\bibinfo  {journal} {Laser \&
  Photonics Reviews}\ } (\bibinfo {year} {2016}),\
  10.1002/lpor.201500321}\BibitemShut {NoStop}%
\bibitem [{\citenamefont {Lodahl}\ \emph {et~al.}(2015)\citenamefont {Lodahl},
  \citenamefont {Mahmoodian},\ and\ \citenamefont {Stobbe}}]{lodahl}%
  \BibitemOpen
  \bibfield  {author} {\bibinfo {author} {\bibfnamefont {P.}~\bibnamefont
  {Lodahl}}, \bibinfo {author} {\bibfnamefont {S.}~\bibnamefont {Mahmoodian}},
  \ and\ \bibinfo {author} {\bibfnamefont {S.}~\bibnamefont {Stobbe}},\
  }\bibfield  {title} {\emph {\bibinfo {title} {Interfacing single photons and
  single quantum dots with photonic nanostructures},\ }}\href {\doibase
  10.1103/RevModPhys.87.347} {\bibfield  {journal} {\bibinfo  {journal} {Rev.
  Mod. Phys.}\ }\textbf {\bibinfo {volume} {87}},\ \bibinfo {pages} {347}
  (\bibinfo {year} {2015})}\BibitemShut {NoStop}%
\bibitem [{\citenamefont {Akahane}\ \emph {et~al.}(2003)\citenamefont
  {Akahane}, \citenamefont {Asano}, \citenamefont {Song},\ and\ \citenamefont
  {Noda}}]{Akahane2003-ma}%
  \BibitemOpen
  \bibfield  {author} {\bibinfo {author} {\bibfnamefont {Y.}~\bibnamefont
  {Akahane}}, \bibinfo {author} {\bibfnamefont {T.}~\bibnamefont {Asano}},
  \bibinfo {author} {\bibfnamefont {B.-S.}\ \bibnamefont {Song}}, \ and\
  \bibinfo {author} {\bibfnamefont {S.}~\bibnamefont {Noda}},\ }\bibfield
  {title} {\emph {\bibinfo {title} {{High-Q} photonic nanocavity in a
  two-dimensional photonic crystal},\ }}\href@noop {} {\bibfield  {journal}
  {\bibinfo  {journal} {Nature}\ }\textbf {\bibinfo {volume} {425}},\ \bibinfo
  {pages} {944} (\bibinfo {year} {2003})}\BibitemShut {NoStop}%
\bibitem [{\citenamefont {Waks}\ \emph {et~al.}(2007)\citenamefont {Waks},
  \citenamefont {Englund}, \citenamefont {Fushman},\ and\ \citenamefont
  {Vu\v{c}kovi\'{c}}}]{Waks2007-lu}%
  \BibitemOpen
  \bibfield  {author} {\bibinfo {author} {\bibfnamefont {E.}~\bibnamefont
  {Waks}}, \bibinfo {author} {\bibfnamefont {D.}~\bibnamefont {Englund}},
  \bibinfo {author} {\bibfnamefont {I.}~\bibnamefont {Fushman}}, \ and\
  \bibinfo {author} {\bibfnamefont {J.}~\bibnamefont {Vu\v{c}kovi\'{c}}},\
  }\bibfield  {title} {\emph {\bibinfo {title} {Efficient photonic crystal
  cavity-waveguide couplers},\ }}\href@noop {} {\bibfield  {journal} {\bibinfo
  {journal} {J. Phys. D Appl. Phys.}\ }\textbf {\bibinfo {volume} {90}},\
  \bibinfo {eid} {073102} (\bibinfo {year} {2007})}\BibitemShut {NoStop}%
\bibitem [{\citenamefont {Yoshie}\ \emph {et~al.}(2004)\citenamefont {Yoshie},
  \citenamefont {Scherer}, \citenamefont {Hendrickson}, \citenamefont
  {Khitrova}, \citenamefont {Gibbs}, \citenamefont {Rupper}, \citenamefont
  {Ell}, \citenamefont {Shchekin},\ and\ \citenamefont
  {Deppe}}]{Yoshie2004-yd}%
  \BibitemOpen
  \bibfield  {author} {\bibinfo {author} {\bibfnamefont {T.}~\bibnamefont
  {Yoshie}}, \bibinfo {author} {\bibfnamefont {A.}~\bibnamefont {Scherer}},
  \bibinfo {author} {\bibfnamefont {J.}~\bibnamefont {Hendrickson}}, \bibinfo
  {author} {\bibfnamefont {G.}~\bibnamefont {Khitrova}}, \bibinfo {author}
  {\bibfnamefont {H.~M.}\ \bibnamefont {Gibbs}}, \bibinfo {author}
  {\bibfnamefont {G.}~\bibnamefont {Rupper}}, \bibinfo {author} {\bibfnamefont
  {C.}~\bibnamefont {Ell}}, \bibinfo {author} {\bibfnamefont {O.~B.}\
  \bibnamefont {Shchekin}}, \ and\ \bibinfo {author} {\bibfnamefont {D.~G.}\
  \bibnamefont {Deppe}},\ }\bibfield  {title} {\emph {\bibinfo {title} {Vacuum
  rabi splitting with a single quantum dot in a photonic crystal nanocavity},\
  }}\href@noop {} {\bibfield  {journal} {\bibinfo  {journal} {Nature}\ }\textbf
  {\bibinfo {volume} {432}},\ \bibinfo {pages} {200} (\bibinfo {year}
  {2004})}\BibitemShut {NoStop}%
\bibitem [{\citenamefont {Englund}\ \emph {et~al.}(2007)\citenamefont
  {Englund}, \citenamefont {Faraon}, \citenamefont {Fushman}, \citenamefont
  {Stoltz}, \citenamefont {Petroff},\ and\ \citenamefont
  {Vuckovi\'{c}}}]{Englund2007-wy}%
  \BibitemOpen
  \bibfield  {author} {\bibinfo {author} {\bibfnamefont {D.}~\bibnamefont
  {Englund}}, \bibinfo {author} {\bibfnamefont {A.}~\bibnamefont {Faraon}},
  \bibinfo {author} {\bibfnamefont {I.}~\bibnamefont {Fushman}}, \bibinfo
  {author} {\bibfnamefont {N.}~\bibnamefont {Stoltz}}, \bibinfo {author}
  {\bibfnamefont {P.}~\bibnamefont {Petroff}}, \ and\ \bibinfo {author}
  {\bibfnamefont {J.}~\bibnamefont {Vuckovi\'{c}}},\ }\bibfield  {title} {\emph
  {\bibinfo {title} {Controlling cavity reflectivity with a single quantum
  dot},\ }}\href@noop {} {\bibfield  {journal} {\bibinfo  {journal} {Nature}\
  }\textbf {\bibinfo {volume} {450}},\ \bibinfo {pages} {857} (\bibinfo {year}
  {2007})}\BibitemShut {NoStop}%
\bibitem [{\citenamefont {Reithmaier}\ \emph {et~al.}(2004)\citenamefont
  {Reithmaier}, \citenamefont {Sek}, \citenamefont {L{\"{o}}ffler},
  \citenamefont {Hofmann}, \citenamefont {Kuhn}, \citenamefont {Reitzenstein},
  \citenamefont {Keldysh}, \citenamefont {Kulakovskii}, \citenamefont
  {Reinecke},\ and\ \citenamefont {Forchel}}]{Reithmaier2004-fu}%
  \BibitemOpen
  \bibfield  {author} {\bibinfo {author} {\bibfnamefont {J.~P.}\ \bibnamefont
  {Reithmaier}}, \bibinfo {author} {\bibfnamefont {G.}~\bibnamefont {Sek}},
  \bibinfo {author} {\bibfnamefont {A.}~\bibnamefont {L{\"{o}}ffler}}, \bibinfo
  {author} {\bibfnamefont {C.}~\bibnamefont {Hofmann}}, \bibinfo {author}
  {\bibfnamefont {S.}~\bibnamefont {Kuhn}}, \bibinfo {author} {\bibfnamefont
  {S.}~\bibnamefont {Reitzenstein}}, \bibinfo {author} {\bibfnamefont {L.~V.}\
  \bibnamefont {Keldysh}}, \bibinfo {author} {\bibfnamefont {V.~D.}\
  \bibnamefont {Kulakovskii}}, \bibinfo {author} {\bibfnamefont {T.~L.}\
  \bibnamefont {Reinecke}}, \ and\ \bibinfo {author} {\bibfnamefont
  {A.}~\bibnamefont {Forchel}},\ }\bibfield  {title} {\emph {\bibinfo {title}
  {Strong coupling in a single quantum dot-semiconductor microcavity system},\
  }}\href@noop {} {\bibfield  {journal} {\bibinfo  {journal} {Nature}\ }\textbf
  {\bibinfo {volume} {432}},\ \bibinfo {pages} {197} (\bibinfo {year}
  {2004})}\BibitemShut {NoStop}%
\bibitem [{\citenamefont {Shore}\ and\ \citenamefont
  {Knight}(1993)}]{Shore1993-zy}%
  \BibitemOpen
  \bibfield  {author} {\bibinfo {author} {\bibfnamefont {B.~W.}\ \bibnamefont
  {Shore}}\ and\ \bibinfo {author} {\bibfnamefont {P.~L.}\ \bibnamefont
  {Knight}},\ }\bibfield  {title} {\emph {\bibinfo {title} {The
  {Jaynes--Cummings} model},\ }}\href {\doibase 10.1080/09500349314551321}
  {\bibfield  {journal} {\bibinfo  {journal} {J. Mod. Opt.}\ }\textbf {\bibinfo
  {volume} {40}},\ \bibinfo {pages} {1195} (\bibinfo {year} {1993})},\ \Eprint
  {http://arxiv.org/abs/http://dx.doi.org/10.1080/09500349314551321}
  {http://dx.doi.org/10.1080/09500349314551321} \BibitemShut {NoStop}%
\bibitem [{\citenamefont {Laussy}\ \emph {et~al.}(2012)\citenamefont {Laussy},
  \citenamefont {del Valle}, \citenamefont {Schrapp}, \citenamefont {Laucht},\
  and\ \citenamefont {Finley}}]{Laussy2012-hr}%
  \BibitemOpen
  \bibfield  {author} {\bibinfo {author} {\bibfnamefont {F.~P.}\ \bibnamefont
  {Laussy}}, \bibinfo {author} {\bibfnamefont {E.}~\bibnamefont {del Valle}},
  \bibinfo {author} {\bibfnamefont {M.}~\bibnamefont {Schrapp}}, \bibinfo
  {author} {\bibfnamefont {A.}~\bibnamefont {Laucht}}, \ and\ \bibinfo {author}
  {\bibfnamefont {J.~J.}\ \bibnamefont {Finley}},\ }\bibfield  {title} {\emph
  {\bibinfo {title} {Climbing the {Jaynes--Cummings} ladder by photon
  counting},\ }}\href@noop {} {\bibfield  {journal} {\bibinfo  {journal} {J.
  Nanophotonics}\ }\textbf {\bibinfo {volume} {6}},\ \bibinfo {pages} {061803}
  (\bibinfo {year} {2012})}\BibitemShut {NoStop}%
\bibitem [{\citenamefont {M{\"{u}}ller}\ \emph
  {et~al.}(2015{\natexlab{a}})\citenamefont {M{\"{u}}ller}, \citenamefont
  {Rundquist}, \citenamefont {Fischer}, \citenamefont {Sarmiento},
  \citenamefont {Lagoudakis}, \citenamefont {Kelaita}, \citenamefont
  {S\'{a}nchez Mu\~{n}oz}, \citenamefont {Del~Valle}, \citenamefont {Laussy},\
  and\ \citenamefont {Vu\v{c}kovi\'{c}}}]{Muller2015-il}%
  \BibitemOpen
  \bibfield  {author} {\bibinfo {author} {\bibfnamefont {K.}~\bibnamefont
  {M{\"{u}}ller}}, \bibinfo {author} {\bibfnamefont {A.}~\bibnamefont
  {Rundquist}}, \bibinfo {author} {\bibfnamefont {K.~A.}\ \bibnamefont
  {Fischer}}, \bibinfo {author} {\bibfnamefont {T.}~\bibnamefont {Sarmiento}},
  \bibinfo {author} {\bibfnamefont {K.~G.}\ \bibnamefont {Lagoudakis}},
  \bibinfo {author} {\bibfnamefont {Y.~A.}\ \bibnamefont {Kelaita}}, \bibinfo
  {author} {\bibfnamefont {C.}~\bibnamefont {S\'{a}nchez Mu\~{n}oz}}, \bibinfo
  {author} {\bibfnamefont {E.}~\bibnamefont {Del~Valle}}, \bibinfo {author}
  {\bibfnamefont {F.~P.}\ \bibnamefont {Laussy}}, \ and\ \bibinfo {author}
  {\bibfnamefont {J.}~\bibnamefont {Vu\v{c}kovi\'{c}}},\ }\bibfield  {title}
  {\emph {\bibinfo {title} {Coherent generation of nonclassical light on chip
  via detuned photon blockade},\ }}\href@noop {} {\bibfield  {journal}
  {\bibinfo  {journal} {Phys. Rev. Lett.}\ }\textbf {\bibinfo {volume} {114}},\
  \bibinfo {pages} {233601} (\bibinfo {year} {2015}{\natexlab{a}})}\BibitemShut
  {NoStop}%
\bibitem [{\citenamefont {M{\"{u}}ller}\ \emph
  {et~al.}(2015{\natexlab{b}})\citenamefont {M{\"{u}}ller}, \citenamefont
  {Fischer}, \citenamefont {Rundquist}, \citenamefont {Dory}, \citenamefont
  {Lagoudakis}, \citenamefont {Sarmiento}, \citenamefont {Kelaita},
  \citenamefont {Borish},\ and\ \citenamefont
  {Vu\v{c}kovi\'{c}}}]{Muller2015-om}%
  \BibitemOpen
  \bibfield  {author} {\bibinfo {author} {\bibfnamefont {K.}~\bibnamefont
  {M{\"{u}}ller}}, \bibinfo {author} {\bibfnamefont {K.~A.}\ \bibnamefont
  {Fischer}}, \bibinfo {author} {\bibfnamefont {A.}~\bibnamefont {Rundquist}},
  \bibinfo {author} {\bibfnamefont {C.}~\bibnamefont {Dory}}, \bibinfo {author}
  {\bibfnamefont {K.~G.}\ \bibnamefont {Lagoudakis}}, \bibinfo {author}
  {\bibfnamefont {T.}~\bibnamefont {Sarmiento}}, \bibinfo {author}
  {\bibfnamefont {Y.~A.}\ \bibnamefont {Kelaita}}, \bibinfo {author}
  {\bibfnamefont {V.}~\bibnamefont {Borish}}, \ and\ \bibinfo {author}
  {\bibfnamefont {J.}~\bibnamefont {Vu\v{c}kovi\'{c}}},\ }\bibfield  {title}
  {\emph {\bibinfo {title} {Ultrafast {Polariton-Phonon} dynamics of strongly
  coupled quantum {Dot-Nanocavity} systems},\ }}\href@noop {} {\bibfield
  {journal} {\bibinfo  {journal} {Phys. Rev. X}\ }\textbf {\bibinfo {volume}
  {5}},\ \bibinfo {pages} {031006} (\bibinfo {year}
  {2015}{\natexlab{b}})}\BibitemShut {NoStop}%
\bibitem [{\citenamefont {M{\"{u}}ller}\ \emph {et~al.}(2016)\citenamefont
  {M{\"{u}}ller}, \citenamefont {Fischer}, \citenamefont {Dory}, \citenamefont
  {Sarmiento}, \citenamefont {Lagoudakis}, \citenamefont {Rundquist},
  \citenamefont {Kelaita},\ and\ \citenamefont
  {Vu\v{c}kovi\'{c}}}]{Muller2016-fs}%
  \BibitemOpen
  \bibfield  {author} {\bibinfo {author} {\bibfnamefont {K.}~\bibnamefont
  {M{\"{u}}ller}}, \bibinfo {author} {\bibfnamefont {K.~A.}\ \bibnamefont
  {Fischer}}, \bibinfo {author} {\bibfnamefont {C.}~\bibnamefont {Dory}},
  \bibinfo {author} {\bibfnamefont {T.}~\bibnamefont {Sarmiento}}, \bibinfo
  {author} {\bibfnamefont {K.~G.}\ \bibnamefont {Lagoudakis}}, \bibinfo
  {author} {\bibfnamefont {A.}~\bibnamefont {Rundquist}}, \bibinfo {author}
  {\bibfnamefont {Y.~A.}\ \bibnamefont {Kelaita}}, \ and\ \bibinfo {author}
  {\bibfnamefont {J.}~\bibnamefont {Vu\v{c}kovi\'{c}}},\ }\bibfield  {title}
  {\emph {\bibinfo {title} {Self-homodyne-enabled generation of
  indistinguishable photons},\ }}\href@noop {} {\bibfield  {journal} {\bibinfo
  {journal} {Optica}\ }\textbf {\bibinfo {volume} {3}},\ \bibinfo {pages} {931}
  (\bibinfo {year} {2016})}\BibitemShut {NoStop}%
\bibitem [{\citenamefont {S\'{a}nchez Mu\~{n}oz}\ \emph
  {et~al.}(2014)\citenamefont {S\'{a}nchez Mu\~{n}oz}, \citenamefont {del
  Valle}, \citenamefont {Gonz\'{a}lez~Tudela}, \citenamefont {M{\"{u}}ller},
  \citenamefont {Lichtmannecker}, \citenamefont {Kaniber}, \citenamefont
  {Tejedor}, \citenamefont {Finley},\ and\ \citenamefont
  {Laussy}}]{Sanchez_Munoz2014-zk}%
  \BibitemOpen
  \bibfield  {author} {\bibinfo {author} {\bibfnamefont {C.}~\bibnamefont
  {S\'{a}nchez Mu\~{n}oz}}, \bibinfo {author} {\bibfnamefont {E.}~\bibnamefont
  {del Valle}}, \bibinfo {author} {\bibfnamefont {A.}~\bibnamefont
  {Gonz\'{a}lez~Tudela}}, \bibinfo {author} {\bibfnamefont {K.}~\bibnamefont
  {M{\"{u}}ller}}, \bibinfo {author} {\bibfnamefont {S.}~\bibnamefont
  {Lichtmannecker}}, \bibinfo {author} {\bibfnamefont {M.}~\bibnamefont
  {Kaniber}}, \bibinfo {author} {\bibfnamefont {C.}~\bibnamefont {Tejedor}},
  \bibinfo {author} {\bibfnamefont {J.~J.}\ \bibnamefont {Finley}}, \ and\
  \bibinfo {author} {\bibfnamefont {F.~P.}\ \bibnamefont {Laussy}},\ }\bibfield
   {title} {\emph {\bibinfo {title} {Emitters of n-photon bundles},\
  }}\href@noop {} {\bibfield  {journal} {\bibinfo  {journal} {Nat. Photonics}\
  }\textbf {\bibinfo {volume} {8}},\ \bibinfo {pages} {550} (\bibinfo {year}
  {2014})}\BibitemShut {NoStop}%
\bibitem [{\citenamefont {Dory}\ \emph
  {et~al.}(2016{\natexlab{a}})\citenamefont {Dory}, \citenamefont {Fischer},
  \citenamefont {M{\"{u}}ller}, \citenamefont {Lagoudakis}, \citenamefont
  {Sarmiento}, \citenamefont {Rundquist}, \citenamefont {Zhang}, \citenamefont
  {Kelaita}, \citenamefont {Sapra},\ and\ \citenamefont
  {Vu\v{c}kovi\'{c}}}]{Dory2016-ca}%
  \BibitemOpen
  \bibfield  {author} {\bibinfo {author} {\bibfnamefont {C.}~\bibnamefont
  {Dory}}, \bibinfo {author} {\bibfnamefont {K.~A.}\ \bibnamefont {Fischer}},
  \bibinfo {author} {\bibfnamefont {K.}~\bibnamefont {M{\"{u}}ller}}, \bibinfo
  {author} {\bibfnamefont {K.~G.}\ \bibnamefont {Lagoudakis}}, \bibinfo
  {author} {\bibfnamefont {T.}~\bibnamefont {Sarmiento}}, \bibinfo {author}
  {\bibfnamefont {A.}~\bibnamefont {Rundquist}}, \bibinfo {author}
  {\bibfnamefont {J.~L.}\ \bibnamefont {Zhang}}, \bibinfo {author}
  {\bibfnamefont {Y.}~\bibnamefont {Kelaita}}, \bibinfo {author} {\bibfnamefont
  {N.~V.}\ \bibnamefont {Sapra}}, \ and\ \bibinfo {author} {\bibfnamefont
  {J.}~\bibnamefont {Vu\v{c}kovi\'{c}}},\ }\bibfield  {title} {\emph {\bibinfo
  {title} {Tuning the photon statistics of a strongly coupled nanophotonic
  system},\ }}\href@noop {} {\  (\bibinfo {year} {2016}{\natexlab{a}})},\
  \Eprint {http://arxiv.org/abs/1610.00174} {arXiv:1610.00174 [physics.optics]}
  \BibitemShut {NoStop}%
\bibitem [{\citenamefont {Fischer}\ \emph {et~al.}(2016)\citenamefont
  {Fischer}, \citenamefont {M{\"{u}}ller}, \citenamefont {Rundquist},
  \citenamefont {Sarmiento}, \citenamefont {Piggott}, \citenamefont {Kelaita},
  \citenamefont {Dory}, \citenamefont {Lagoudakis},\ and\ \citenamefont
  {Vu\v{c}kovi\'{c}}}]{Fischer2016-dj}%
  \BibitemOpen
  \bibfield  {author} {\bibinfo {author} {\bibfnamefont {K.~A.}\ \bibnamefont
  {Fischer}}, \bibinfo {author} {\bibfnamefont {K.}~\bibnamefont
  {M{\"{u}}ller}}, \bibinfo {author} {\bibfnamefont {A.}~\bibnamefont
  {Rundquist}}, \bibinfo {author} {\bibfnamefont {T.}~\bibnamefont
  {Sarmiento}}, \bibinfo {author} {\bibfnamefont {A.~Y.}\ \bibnamefont
  {Piggott}}, \bibinfo {author} {\bibfnamefont {Y.}~\bibnamefont {Kelaita}},
  \bibinfo {author} {\bibfnamefont {C.}~\bibnamefont {Dory}}, \bibinfo {author}
  {\bibfnamefont {K.~G.}\ \bibnamefont {Lagoudakis}}, \ and\ \bibinfo {author}
  {\bibfnamefont {J.}~\bibnamefont {Vu\v{c}kovi\'{c}}},\ }\bibfield  {title}
  {\emph {\bibinfo {title} {Self-homodyne measurement of a dynamic mollow
  triplet in the solid state},\ }}\href@noop {} {\bibfield  {journal} {\bibinfo
   {journal} {Nat. Photonics}\ }\textbf {\bibinfo {volume} {10}},\ \bibinfo
  {pages} {163} (\bibinfo {year} {2016})}\BibitemShut {NoStop}%
\bibitem [{\citenamefont {Johansson}\ \emph {et~al.}(2013)\citenamefont
  {Johansson}, \citenamefont {Nation},\ and\ \citenamefont
  {Nori}}]{Johansson2013-cv}%
  \BibitemOpen
  \bibfield  {author} {\bibinfo {author} {\bibfnamefont {J.~R.}\ \bibnamefont
  {Johansson}}, \bibinfo {author} {\bibfnamefont {P.~D.}\ \bibnamefont
  {Nation}}, \ and\ \bibinfo {author} {\bibfnamefont {F.}~\bibnamefont
  {Nori}},\ }\bibfield  {title} {\emph {\bibinfo {title} {{QuTiP} 2: A python
  framework for the dynamics of open quantum systems},\ }}\href@noop {}
  {\bibfield  {journal} {\bibinfo  {journal} {Comput. Phys. Commun.}\ }\textbf
  {\bibinfo {volume} {184}},\ \bibinfo {pages} {1234} (\bibinfo {year}
  {2013})}\BibitemShut {NoStop}%
\bibitem [{\citenamefont {Yu}\ \emph {et~al.}(2014)\citenamefont {Yu},
  \citenamefont {Heuck}, \citenamefont {Hu}, \citenamefont {Xue}, \citenamefont
  {Peucheret}, \citenamefont {Chen}, \citenamefont {Oxenl\o{}we}, \citenamefont
  {Yvind},\ and\ \citenamefont {M\o{}rk}}]{Yu2014-qp}%
  \BibitemOpen
  \bibfield  {author} {\bibinfo {author} {\bibfnamefont {Y.}~\bibnamefont
  {Yu}}, \bibinfo {author} {\bibfnamefont {M.}~\bibnamefont {Heuck}}, \bibinfo
  {author} {\bibfnamefont {H.}~\bibnamefont {Hu}}, \bibinfo {author}
  {\bibfnamefont {W.}~\bibnamefont {Xue}}, \bibinfo {author} {\bibfnamefont
  {C.}~\bibnamefont {Peucheret}}, \bibinfo {author} {\bibfnamefont
  {Y.}~\bibnamefont {Chen}}, \bibinfo {author} {\bibfnamefont {L.~K.}\
  \bibnamefont {Oxenl\o{}we}}, \bibinfo {author} {\bibfnamefont
  {K.}~\bibnamefont {Yvind}}, \ and\ \bibinfo {author} {\bibfnamefont
  {J.}~\bibnamefont {M\o{}rk}},\ }\bibfield  {title} {\emph {\bibinfo {title}
  {Fano resonance control in a photonic crystal structure and its application
  to ultrafast switching},\ }}\href@noop {} {\bibfield  {journal} {\bibinfo
  {journal} {Appl. Phys. Lett.}\ }\textbf {\bibinfo {volume} {105}},\ \bibinfo
  {pages} {061117} (\bibinfo {year} {2014})}\BibitemShut {NoStop}%
\bibitem [{\citenamefont {Fano}(1961)}]{Fano1961-oa}%
  \BibitemOpen
  \bibfield  {author} {\bibinfo {author} {\bibfnamefont {U.}~\bibnamefont
  {Fano}},\ }\bibfield  {title} {\emph {\bibinfo {title} {Effects of
  configuration interaction on intensities and phase shifts},\ }}\href@noop {}
  {\bibfield  {journal} {\bibinfo  {journal} {Phys. Rev.}\ }\textbf {\bibinfo
  {volume} {124}},\ \bibinfo {pages} {1866} (\bibinfo {year}
  {1961})}\BibitemShut {NoStop}%
\bibitem [{\citenamefont {Yu}\ \emph {et~al.}(2016{\natexlab{a}})\citenamefont
  {Yu}, \citenamefont {Xue}, \citenamefont {Hu}, \citenamefont {Oxenl\o{}we},
  \citenamefont {Yvind},\ and\ \citenamefont {Mork}}]{Yu2016-tn}%
  \BibitemOpen
  \bibfield  {author} {\bibinfo {author} {\bibfnamefont {Y.}~\bibnamefont
  {Yu}}, \bibinfo {author} {\bibfnamefont {W.}~\bibnamefont {Xue}}, \bibinfo
  {author} {\bibfnamefont {H.}~\bibnamefont {Hu}}, \bibinfo {author}
  {\bibfnamefont {L.~K.}\ \bibnamefont {Oxenl\o{}we}}, \bibinfo {author}
  {\bibfnamefont {K.}~\bibnamefont {Yvind}}, \ and\ \bibinfo {author}
  {\bibfnamefont {J.}~\bibnamefont {Mork}},\ }\bibfield  {title} {\emph
  {\bibinfo {title} {{All-Optical} switching improvement using
  {Photonic-Crystal} fano structures},\ }}\href@noop {} {\bibfield  {journal}
  {\bibinfo  {journal} {IEEE Photonics J.}\ }\textbf {\bibinfo {volume} {8}},\
  \bibinfo {pages} {1} (\bibinfo {year} {2016}{\natexlab{a}})}\BibitemShut
  {NoStop}%
\bibitem [{\citenamefont {Yu}\ \emph {et~al.}(2015)\citenamefont {Yu},
  \citenamefont {Hu}, \citenamefont {Oxenl\o{}we}, \citenamefont {Yvind},\ and\
  \citenamefont {Mork}}]{Yu2015-mh}%
  \BibitemOpen
  \bibfield  {author} {\bibinfo {author} {\bibfnamefont {Y.}~\bibnamefont
  {Yu}}, \bibinfo {author} {\bibfnamefont {H.}~\bibnamefont {Hu}}, \bibinfo
  {author} {\bibfnamefont {L.~K.}\ \bibnamefont {Oxenl\o{}we}}, \bibinfo
  {author} {\bibfnamefont {K.}~\bibnamefont {Yvind}}, \ and\ \bibinfo {author}
  {\bibfnamefont {J.}~\bibnamefont {Mork}},\ }\bibfield  {title} {\emph
  {\bibinfo {title} {Ultrafast all-optical modulation using a photonic-crystal
  fano structure with broken symmetry},\ }}\href@noop {} {\bibfield  {journal}
  {\bibinfo  {journal} {Opt. Lett.}\ }\textbf {\bibinfo {volume} {40}},\
  \bibinfo {pages} {2357} (\bibinfo {year} {2015})}\BibitemShut {NoStop}%
\bibitem [{\citenamefont {Zhao}\ \emph {et~al.}(2016)\citenamefont {Zhao},
  \citenamefont {Qian}, \citenamefont {Qiu}, \citenamefont {Tang},
  \citenamefont {Sun}, \citenamefont {Jin},\ and\ \citenamefont
  {Xu}}]{Zhao2016-qs}%
  \BibitemOpen
  \bibfield  {author} {\bibinfo {author} {\bibfnamefont {Y.}~\bibnamefont
  {Zhao}}, \bibinfo {author} {\bibfnamefont {C.}~\bibnamefont {Qian}}, \bibinfo
  {author} {\bibfnamefont {K.}~\bibnamefont {Qiu}}, \bibinfo {author}
  {\bibfnamefont {J.}~\bibnamefont {Tang}}, \bibinfo {author} {\bibfnamefont
  {Y.}~\bibnamefont {Sun}}, \bibinfo {author} {\bibfnamefont {K.}~\bibnamefont
  {Jin}}, \ and\ \bibinfo {author} {\bibfnamefont {X.}~\bibnamefont {Xu}},\
  }\bibfield  {title} {\emph {\bibinfo {title} {Gain enhanced fano resonance in
  a coupled photonic crystal cavity-waveguide structure},\ }}\href@noop {}
  {\bibfield  {journal} {\bibinfo  {journal} {Sci. Rep.}\ }\textbf {\bibinfo
  {volume} {6}},\ \bibinfo {pages} {33645} (\bibinfo {year}
  {2016})}\BibitemShut {NoStop}%
\bibitem [{\citenamefont {Li}\ \emph {et~al.}(2016)\citenamefont {Li},
  \citenamefont {Yu}, \citenamefont {Liu}, \citenamefont {Ding},\ and\
  \citenamefont {Wu}}]{Li2016-sm}%
  \BibitemOpen
  \bibfield  {author} {\bibinfo {author} {\bibfnamefont {J.}~\bibnamefont
  {Li}}, \bibinfo {author} {\bibfnamefont {R.}~\bibnamefont {Yu}}, \bibinfo
  {author} {\bibfnamefont {J.}~\bibnamefont {Liu}}, \bibinfo {author}
  {\bibfnamefont {C.}~\bibnamefont {Ding}}, \ and\ \bibinfo {author}
  {\bibfnamefont {Y.}~\bibnamefont {Wu}},\ }\bibfield  {title} {\emph {\bibinfo
  {title} {Fano line-shape control and superluminal light using cavity quantum
  electrodynamics with a partially transmitting element},\ }}\href@noop {}
  {\bibfield  {journal} {\bibinfo  {journal} {Phys. Rev. A}\ }\textbf {\bibinfo
  {volume} {93}},\ \bibinfo {pages} {053814} (\bibinfo {year}
  {2016})}\BibitemShut {NoStop}%
\bibitem [{\citenamefont {Yu}\ \emph {et~al.}(2016{\natexlab{b}})\citenamefont
  {Yu}, \citenamefont {Xue}, \citenamefont {Semenova}, \citenamefont {Yvind},\
  and\ \citenamefont {Mork}}]{Yu2016-me}%
  \BibitemOpen
  \bibfield  {author} {\bibinfo {author} {\bibfnamefont {Y.}~\bibnamefont
  {Yu}}, \bibinfo {author} {\bibfnamefont {W.}~\bibnamefont {Xue}}, \bibinfo
  {author} {\bibfnamefont {E.}~\bibnamefont {Semenova}}, \bibinfo {author}
  {\bibfnamefont {K.}~\bibnamefont {Yvind}}, \ and\ \bibinfo {author}
  {\bibfnamefont {J.}~\bibnamefont {Mork}},\ }\bibfield  {title} {\emph
  {\bibinfo {title} {Demonstration of a self-pulsing photonic crystal fano
  laser},\ }}\href@noop {} {\  (\bibinfo {year} {2016}{\natexlab{b}})},\
  \Eprint {http://arxiv.org/abs/1605.03028} {arXiv:1605.03028 [physics.optics]}
  \BibitemShut {NoStop}%
\bibitem [{\citenamefont {Haroche}\ and\ \citenamefont
  {Raimond}(2006)}]{Haroche2006-cq}%
  \BibitemOpen
  \bibfield  {author} {\bibinfo {author} {\bibfnamefont {S.}~\bibnamefont
  {Haroche}}\ and\ \bibinfo {author} {\bibfnamefont {J.~M.}\ \bibnamefont
  {Raimond}},\ }\href@noop {} {\emph {\bibinfo {title} {Exploring the Quantum:
  Atoms, Cavities, and Photons}}},\ Oxford Graduate Texts\ (\bibinfo
  {publisher} {OUP Oxford},\ \bibinfo {year} {2006})\BibitemShut {NoStop}%
\bibitem [{\citenamefont {Loudon}(2000)}]{Loudon2000-li}%
  \BibitemOpen
  \bibfield  {author} {\bibinfo {author} {\bibfnamefont {R.}~\bibnamefont
  {Loudon}},\ }\href@noop {} {\emph {\bibinfo {title} {The Quantum Theory of
  Light}}}\ (\bibinfo  {publisher} {OUP Oxford},\ \bibinfo {year}
  {2000})\BibitemShut {NoStop}%
\bibitem [{\citenamefont {Joannopoulos}\ \emph {et~al.}(2011)\citenamefont
  {Joannopoulos}, \citenamefont {Johnson}, \citenamefont {Winn},\ and\
  \citenamefont {Meade}}]{Joannopoulos2011-ax}%
  \BibitemOpen
  \bibfield  {author} {\bibinfo {author} {\bibfnamefont {J.~D.}\ \bibnamefont
  {Joannopoulos}}, \bibinfo {author} {\bibfnamefont {S.~G.}\ \bibnamefont
  {Johnson}}, \bibinfo {author} {\bibfnamefont {J.~N.}\ \bibnamefont {Winn}}, \
  and\ \bibinfo {author} {\bibfnamefont {R.~D.}\ \bibnamefont {Meade}},\
  }\bibfield  {title} {\emph {\bibinfo {title} {Photonic crystals: molding the
  flow of light},\ }}\href {http://ab-initio.mit.edu/book/} {\  (\bibinfo
  {year} {2011})}\BibitemShut {NoStop}%
\bibitem [{\citenamefont {Faraon}\ \emph {et~al.}(2008)\citenamefont {Faraon},
  \citenamefont {Fushman}, \citenamefont {Englund}, \citenamefont {Stoltz},
  \citenamefont {Petroff},\ and\ \citenamefont
  {Vu\v{c}kovi\'{c}}}]{Faraon2008-zh}%
  \BibitemOpen
  \bibfield  {author} {\bibinfo {author} {\bibfnamefont {A.}~\bibnamefont
  {Faraon}}, \bibinfo {author} {\bibfnamefont {I.}~\bibnamefont {Fushman}},
  \bibinfo {author} {\bibfnamefont {D.}~\bibnamefont {Englund}}, \bibinfo
  {author} {\bibfnamefont {N.}~\bibnamefont {Stoltz}}, \bibinfo {author}
  {\bibfnamefont {P.}~\bibnamefont {Petroff}}, \ and\ \bibinfo {author}
  {\bibfnamefont {J.}~\bibnamefont {Vu\v{c}kovi\'{c}}},\ }\bibfield  {title}
  {\emph {\bibinfo {title} {Coherent generation of non-classical light on a
  chip via photon-induced tunnelling and blockade},\ }}\href@noop {} {\bibfield
   {journal} {\bibinfo  {journal} {Nat. Phys.}\ }\textbf {\bibinfo {volume}
  {4}},\ \bibinfo {pages} {859} (\bibinfo {year} {2008})}\BibitemShut {NoStop}%
\bibitem [{\citenamefont {Reinhard}\ \emph {et~al.}(2011)\citenamefont
  {Reinhard}, \citenamefont {Volz}, \citenamefont {Winger}, \citenamefont
  {Badolato}, \citenamefont {Hennessy}, \citenamefont {Hu},\ and\ \citenamefont
  {Imamo\u{g}lu}}]{Reinhard2011-ye}%
  \BibitemOpen
  \bibfield  {author} {\bibinfo {author} {\bibfnamefont {A.}~\bibnamefont
  {Reinhard}}, \bibinfo {author} {\bibfnamefont {T.}~\bibnamefont {Volz}},
  \bibinfo {author} {\bibfnamefont {M.}~\bibnamefont {Winger}}, \bibinfo
  {author} {\bibfnamefont {A.}~\bibnamefont {Badolato}}, \bibinfo {author}
  {\bibfnamefont {K.~J.}\ \bibnamefont {Hennessy}}, \bibinfo {author}
  {\bibfnamefont {E.~L.}\ \bibnamefont {Hu}}, \ and\ \bibinfo {author}
  {\bibfnamefont {A.}~\bibnamefont {Imamo\u{g}lu}},\ }\bibfield  {title} {\emph
  {\bibinfo {title} {Strongly correlated photons on a chip},\ }}\href@noop {}
  {\bibfield  {journal} {\bibinfo  {journal} {Nat. Photonics}\ }\textbf
  {\bibinfo {volume} {6}},\ \bibinfo {pages} {93} (\bibinfo {year}
  {2011})}\BibitemShut {NoStop}%
\bibitem [{\citenamefont {Englund}\ \emph {et~al.}(2005)\citenamefont
  {Englund}, \citenamefont {Fattal}, \citenamefont {Waks}, \citenamefont
  {Solomon}, \citenamefont {Zhang}, \citenamefont {Nakaoka}, \citenamefont
  {Arakawa}, \citenamefont {Yamamoto},\ and\ \citenamefont
  {Vuckovi{\'c}}}]{Englund2005-wz}%
  \BibitemOpen
  \bibfield  {author} {\bibinfo {author} {\bibfnamefont {D.}~\bibnamefont
  {Englund}}, \bibinfo {author} {\bibfnamefont {D.}~\bibnamefont {Fattal}},
  \bibinfo {author} {\bibfnamefont {E.}~\bibnamefont {Waks}}, \bibinfo {author}
  {\bibfnamefont {G.}~\bibnamefont {Solomon}}, \bibinfo {author} {\bibfnamefont
  {B.}~\bibnamefont {Zhang}}, \bibinfo {author} {\bibfnamefont
  {T.}~\bibnamefont {Nakaoka}}, \bibinfo {author} {\bibfnamefont
  {Y.}~\bibnamefont {Arakawa}}, \bibinfo {author} {\bibfnamefont
  {Y.}~\bibnamefont {Yamamoto}}, \ and\ \bibinfo {author} {\bibfnamefont
  {J.}~\bibnamefont {Vuckovi{\'c}}},\ }\bibfield  {title} {\emph {\bibinfo
  {title} {Controlling the spontaneous emission rate of single quantum dots in
  a two-dimensional photonic crystal},\ }}\href@noop {} {\bibfield  {journal}
  {\bibinfo  {journal} {Phys. Rev. Lett.}\ }\textbf {\bibinfo {volume} {95}},\
  \bibinfo {pages} {013904} (\bibinfo {year} {2005})}\BibitemShut {NoStop}%
\bibitem [{\citenamefont {Radulaski}\ \emph {et~al.}(2017)\citenamefont
  {Radulaski}, \citenamefont {Fischer},\ and\ \citenamefont
  {Vuckovic}}]{Radulaski2017-hj}%
  \BibitemOpen
  \bibfield  {author} {\bibinfo {author} {\bibfnamefont {M.}~\bibnamefont
  {Radulaski}}, \bibinfo {author} {\bibfnamefont {K.}~\bibnamefont {Fischer}},
  \ and\ \bibinfo {author} {\bibfnamefont {J.}~\bibnamefont {Vuckovic}},\
  }\bibfield  {title} {\emph {\bibinfo {title} {Nonclassical light generation
  from {III-V} and {Group-IV} {Solid-State} cavity quantum systems},\
  }}\href@noop {} {\  (\bibinfo {year} {2017})},\ \Eprint
  {http://arxiv.org/abs/1701.03039} {arXiv:1701.03039 [physics.optics]}
  \BibitemShut {NoStop}%
\bibitem [{\citenamefont {Zubizarreta~Casalengua}\ \emph
  {et~al.}(2016)\citenamefont {Zubizarreta~Casalengua}, \citenamefont
  {L\'{o}pez Carre\~{n}o}, \citenamefont {del Valle},\ and\ \citenamefont
  {Laussy}}]{Zubizarreta_Casalengua2016-wj}%
  \BibitemOpen
  \bibfield  {author} {\bibinfo {author} {\bibfnamefont {E.}~\bibnamefont
  {Zubizarreta~Casalengua}}, \bibinfo {author} {\bibfnamefont {J.~C.}\
  \bibnamefont {L\'{o}pez Carre\~{n}o}}, \bibinfo {author} {\bibfnamefont
  {E.}~\bibnamefont {del Valle}}, \ and\ \bibinfo {author} {\bibfnamefont
  {F.~P.}\ \bibnamefont {Laussy}},\ }\bibfield  {title} {\emph {\bibinfo
  {title} {Structure of the harmonic oscillator hilbert space},\ }}\href@noop
  {} {\  (\bibinfo {year} {2016})},\ \Eprint {http://arxiv.org/abs/1607.03976}
  {arXiv:1607.03976 [quant-ph]} \BibitemShut {NoStop}%
\bibitem [{\citenamefont {Carre\~{n}o}\ \emph {et~al.}(2016)\citenamefont
  {Carre\~{n}o}, \citenamefont {Casalengua}, \citenamefont {del Valle},\ and\
  \citenamefont {Laussy}}]{Camilo-xn}%
  \BibitemOpen
  \bibfield  {author} {\bibinfo {author} {\bibfnamefont {J.~C.~L.}\
  \bibnamefont {Carre\~{n}o}}, \bibinfo {author} {\bibfnamefont {E.~Z.}\
  \bibnamefont {Casalengua}}, \bibinfo {author} {\bibfnamefont
  {E.}~\bibnamefont {del Valle}}, \ and\ \bibinfo {author} {\bibfnamefont
  {F.~P.}\ \bibnamefont {Laussy}},\ }\bibfield  {title} {\emph {\bibinfo
  {title} {Criterion for single photon sources},\ }}\href@noop {} {\  (\bibinfo
  {year} {2016})},\ \Eprint {http://arxiv.org/abs/1610.06126} {arXiv:1610.06126
  [quant-ph]} \BibitemShut {NoStop}%
\bibitem [{\citenamefont {Gardiner}\ and\ \citenamefont
  {Zoller}(2004)}]{Gardiner2004-qu}%
  \BibitemOpen
  \bibfield  {author} {\bibinfo {author} {\bibfnamefont {C.}~\bibnamefont
  {Gardiner}}\ and\ \bibinfo {author} {\bibfnamefont {P.}~\bibnamefont
  {Zoller}},\ }\href@noop {} {\emph {\bibinfo {title} {Quantum Noise: A
  Handbook of Markovian and {Non-Markovian} Quantum Stochastic Methods with
  Applications to Quantum Optics}}},\ Springer Series in Synergetics\ (\bibinfo
   {publisher} {Springer},\ \bibinfo {year} {2004})\BibitemShut {NoStop}%
\bibitem [{\citenamefont {Carmichael}(2009)}]{Carmichael2009-nj}%
  \BibitemOpen
  \bibfield  {author} {\bibinfo {author} {\bibfnamefont {H.~J.}\ \bibnamefont
  {Carmichael}},\ }\href@noop {} {\emph {\bibinfo {title} {Statistical Methods
  in Quantum Optics}}},\ Theoretical and Mathematical Physics\ (\bibinfo
  {publisher} {Springer Berlin Heidelberg},\ \bibinfo {year}
  {2009})\BibitemShut {NoStop}%
\bibitem [{\citenamefont {Dory}\ \emph
  {et~al.}(2016{\natexlab{b}})\citenamefont {Dory}, \citenamefont {Fischer},
  \citenamefont {M{\"{u}}ller}, \citenamefont {Lagoudakis}, \citenamefont
  {Sarmiento}, \citenamefont {Rundquist}, \citenamefont {Zhang}, \citenamefont
  {Kelaita},\ and\ \citenamefont {Vu\v{c}kovi\'{c}}}]{Dory2016-rf}%
  \BibitemOpen
  \bibfield  {author} {\bibinfo {author} {\bibfnamefont {C.}~\bibnamefont
  {Dory}}, \bibinfo {author} {\bibfnamefont {K.~A.}\ \bibnamefont {Fischer}},
  \bibinfo {author} {\bibfnamefont {K.}~\bibnamefont {M{\"{u}}ller}}, \bibinfo
  {author} {\bibfnamefont {K.~G.}\ \bibnamefont {Lagoudakis}}, \bibinfo
  {author} {\bibfnamefont {T.}~\bibnamefont {Sarmiento}}, \bibinfo {author}
  {\bibfnamefont {A.}~\bibnamefont {Rundquist}}, \bibinfo {author}
  {\bibfnamefont {J.~L.}\ \bibnamefont {Zhang}}, \bibinfo {author}
  {\bibfnamefont {Y.}~\bibnamefont {Kelaita}}, \ and\ \bibinfo {author}
  {\bibfnamefont {J.}~\bibnamefont {Vu\v{c}kovi\'{c}}},\ }\bibfield  {title}
  {\emph {\bibinfo {title} {Complete coherent control of a quantum dot strongly
  coupled to a nanocavity},\ }}\href@noop {} {\bibfield  {journal} {\bibinfo
  {journal} {Sci. Rep.}\ }\textbf {\bibinfo {volume} {6}},\ \bibinfo {pages}
  {25172} (\bibinfo {year} {2016}{\natexlab{b}})}\BibitemShut {NoStop}%
\bibitem [{\citenamefont {Schuster}\ \emph {et~al.}(2008)\citenamefont
  {Schuster}, \citenamefont {Kubanek}, \citenamefont {Fuhrmanek}, \citenamefont
  {Puppe}, \citenamefont {Pinkse}, \citenamefont {Murr},\ and\ \citenamefont
  {Rempe}}]{Schuster2008-go}%
  \BibitemOpen
  \bibfield  {author} {\bibinfo {author} {\bibfnamefont {I.}~\bibnamefont
  {Schuster}}, \bibinfo {author} {\bibfnamefont {A.}~\bibnamefont {Kubanek}},
  \bibinfo {author} {\bibfnamefont {A.}~\bibnamefont {Fuhrmanek}}, \bibinfo
  {author} {\bibfnamefont {T.}~\bibnamefont {Puppe}}, \bibinfo {author}
  {\bibfnamefont {P.~W.~H.}\ \bibnamefont {Pinkse}}, \bibinfo {author}
  {\bibfnamefont {K.}~\bibnamefont {Murr}}, \ and\ \bibinfo {author}
  {\bibfnamefont {G.}~\bibnamefont {Rempe}},\ }\bibfield  {title} {\emph
  {\bibinfo {title} {Nonlinear spectroscopy of photons bound to one atom},\
  }}\href@noop {} {\bibfield  {journal} {\bibinfo  {journal} {Nat. Phys.}\
  }\textbf {\bibinfo {volume} {4}},\ \bibinfo {pages} {382} (\bibinfo {year}
  {2008})}\BibitemShut {NoStop}%
\bibitem [{\citenamefont {Arakawa}\ \emph {et~al.}(2012)\citenamefont
  {Arakawa}, \citenamefont {Iwamoto}, \citenamefont {Nomura}, \citenamefont
  {Tandaechanurat},\ and\ \citenamefont {Ota}}]{Arakawa2012-yd}%
  \BibitemOpen
  \bibfield  {author} {\bibinfo {author} {\bibfnamefont {Y.}~\bibnamefont
  {Arakawa}}, \bibinfo {author} {\bibfnamefont {S.}~\bibnamefont {Iwamoto}},
  \bibinfo {author} {\bibfnamefont {M.}~\bibnamefont {Nomura}}, \bibinfo
  {author} {\bibfnamefont {A.}~\bibnamefont {Tandaechanurat}}, \ and\ \bibinfo
  {author} {\bibfnamefont {Y.}~\bibnamefont {Ota}},\ }\bibfield  {title} {\emph
  {\bibinfo {title} {Cavity quantum electrodynamics and lasing oscillation in
  single quantum dot-photonic crystal nanocavity coupled systems},\
  }}\href@noop {} {\bibfield  {journal} {\bibinfo  {journal} {IEEE Journal of
  Selected Topics in Quantum Electronics}\ }\textbf {\bibinfo {volume} {18}},\
  \bibinfo {pages} {1818} (\bibinfo {year} {2012})}\BibitemShut {NoStop}%
\bibitem [{\citenamefont {Guha}\ \emph {et~al.}(2017)\citenamefont {Guha},
  \citenamefont {Marsault}, \citenamefont {Cadiz}, \citenamefont {Morgenroth},
  \citenamefont {Ulin}, \citenamefont {Berkovitz}, \citenamefont
  {Lema{\^\i}tre}, \citenamefont {Gomez}, \citenamefont {Amo}, \citenamefont
  {Combri{\'e}}, \citenamefont {G{\'e}rard}, \citenamefont {Leo},\ and\
  \citenamefont {Favero}}]{Guha2017-qy}%
  \BibitemOpen
  \bibfield  {author} {\bibinfo {author} {\bibfnamefont {B.}~\bibnamefont
  {Guha}}, \bibinfo {author} {\bibfnamefont {F.}~\bibnamefont {Marsault}},
  \bibinfo {author} {\bibfnamefont {F.}~\bibnamefont {Cadiz}}, \bibinfo
  {author} {\bibfnamefont {L.}~\bibnamefont {Morgenroth}}, \bibinfo {author}
  {\bibfnamefont {V.}~\bibnamefont {Ulin}}, \bibinfo {author} {\bibfnamefont
  {V.}~\bibnamefont {Berkovitz}}, \bibinfo {author} {\bibfnamefont
  {A.}~\bibnamefont {Lema{\^\i}tre}}, \bibinfo {author} {\bibfnamefont
  {C.}~\bibnamefont {Gomez}}, \bibinfo {author} {\bibfnamefont
  {A.}~\bibnamefont {Amo}}, \bibinfo {author} {\bibfnamefont {S.}~\bibnamefont
  {Combri{\'e}}}, \bibinfo {author} {\bibfnamefont {B.}~\bibnamefont
  {G{\'e}rard}}, \bibinfo {author} {\bibfnamefont {G.}~\bibnamefont {Leo}}, \
  and\ \bibinfo {author} {\bibfnamefont {I.}~\bibnamefont {Favero}},\
  }\bibfield  {title} {\emph {\bibinfo {title} {Surface-enhanced gallium
  arsenide photonic resonator with quality factor of 6 $\times$ 10$^{6}$},\
  }}\href@noop {} {\bibfield  {journal} {\bibinfo  {journal} {Optica}\ }\textbf
  {\bibinfo {volume} {4}},\ \bibinfo {pages} {218} (\bibinfo {year}
  {2017})}\BibitemShut {NoStop}%
\bibitem [{\citenamefont {Iles-Smith}\ \emph {et~al.}(2016)\citenamefont
  {Iles-Smith}, \citenamefont {McCutcheon}, \citenamefont {Nazir},\ and\
  \citenamefont {M{\o}rk}}]{Iles-Smith2016-bp}%
  \BibitemOpen
  \bibfield  {author} {\bibinfo {author} {\bibfnamefont {J.}~\bibnamefont
  {Iles-Smith}}, \bibinfo {author} {\bibfnamefont {D.~P.~S.}\ \bibnamefont
  {McCutcheon}}, \bibinfo {author} {\bibfnamefont {A.}~\bibnamefont {Nazir}}, \
  and\ \bibinfo {author} {\bibfnamefont {J.}~\bibnamefont {M{\o}rk}},\
  }\bibfield  {title} {\emph {\bibinfo {title} {Phonon limit to simultaneous
  near-unity efficiency and indistinguishability in semiconductor single photon
  sources},\ }}\href@noop {} {\  (\bibinfo {year} {2016})},\ \Eprint
  {http://arxiv.org/abs/1612.04173} {arXiv:1612.04173 [quant-ph]} \BibitemShut
  {NoStop}%
\bibitem [{\citenamefont {Grange}\ \emph {et~al.}(2016)\citenamefont {Grange},
  \citenamefont {Somaschi}, \citenamefont {Ant{\'o}n}, \citenamefont
  {De~Santis}, \citenamefont {Coppola}, \citenamefont {Giesz}, \citenamefont
  {Lema{\^\i}tre}, \citenamefont {Sagnes}, \citenamefont {Auff{\`e}ves},\ and\
  \citenamefont {Senellart}}]{Grange2016-yp}%
  \BibitemOpen
  \bibfield  {author} {\bibinfo {author} {\bibfnamefont {T.}~\bibnamefont
  {Grange}}, \bibinfo {author} {\bibfnamefont {N.}~\bibnamefont {Somaschi}},
  \bibinfo {author} {\bibfnamefont {C.}~\bibnamefont {Ant{\'o}n}}, \bibinfo
  {author} {\bibfnamefont {L.}~\bibnamefont {De~Santis}}, \bibinfo {author}
  {\bibfnamefont {G.}~\bibnamefont {Coppola}}, \bibinfo {author} {\bibfnamefont
  {V.}~\bibnamefont {Giesz}}, \bibinfo {author} {\bibfnamefont
  {A.}~\bibnamefont {Lema{\^\i}tre}}, \bibinfo {author} {\bibfnamefont
  {I.}~\bibnamefont {Sagnes}}, \bibinfo {author} {\bibfnamefont
  {A.}~\bibnamefont {Auff{\`e}ves}}, \ and\ \bibinfo {author} {\bibfnamefont
  {P.}~\bibnamefont {Senellart}},\ }\bibfield  {title} {\emph {\bibinfo {title}
  {Overcoming phonon-induced decoherence in single-photon sources with cavity
  quantum electrodynamics},\ }}\href@noop {} {\  (\bibinfo {year} {2016})},\
  \Eprint {http://arxiv.org/abs/1612.03063} {arXiv:1612.03063 [quant-ph]}
  \BibitemShut {NoStop}%
\end{thebibliography}%
\end{document}